\documentclass[10pt, conference]{IEEEtran}
\IEEEoverridecommandlockouts
\usepackage{cite}
\usepackage{amsmath,amssymb,amsfonts}
\usepackage{algorithmic}
\usepackage{graphicx}
\usepackage{textcomp}
\usepackage{xcolor}
\usepackage{microtype}
\def\BibTeX{{\rm B\kern-.05em{\sc i\kern-.025em b}\kern-.08em
    T\kern-.1667em\lower.7ex\hbox{E}\kern-.125emX}}

\usepackage{amsfonts}  
\usepackage{mathtools}       % allow usage of \vcentcolon
\usepackage{tikz,pgfplots}
\usetikzlibrary{arrows,decorations.markings}
\usepackage{varwidth}        % used in spread overview plot
\usepackage{ifthen}          % used in spread overview plot
\usepgfplotslibrary{groupplots}
\usepackage{pgfplotstable}   % used in sep timing table
\usepackage{booktabs}        % toprule, etc in table
\usepackage{caption}         % used in sep timing table 
\usepackage{cleveref}  
\usepackage{pythonhighlight}
\usepackage{svg}
\usepackage{url}
\pgfplotsset{
	tick label style = {font ={\large}},
	label style = {font = {\large}},
	legend style = {font = {\large}},
	title style = {font = {\Large}},
}

\newcommand{\defeq}{\vcentcolon=}

\newtheorem{rem}{Remark}

\newcommand{\bi}{\begin{itemize}}
\newcommand{\ei}{\end{itemize}}
\newcommand{\ben}{\begin{enumerate}}
\newcommand{\een}{\end{enumerate}}
\newcommand{\be}{\begin{equation}}
\newcommand{\ee}{\end{equation}}
\newcommand{\bea}{\begin{eqnarray}} 
\newcommand{\eea}{\end{eqnarray}}
\newcommand{\ba}{\begin{align}} 
\newcommand{\ea}{\end{align}}
\newcommand{\bse}{\begin{subequations}} 
\newcommand{\ese}{\end{subequations}}
\newcommand{\bc}{\begin{center}}
\newcommand{\ec}{\end{center}}
\newcommand{\bfi}{\begin{figure}}
\newcommand{\efi}{\end{figure}}

\newcommand{\bmp}[1]{\begin{minipage}{#1}}
\newcommand{\emp}{\end{minipage}}

\newcommand{\ZZ}{\mathbb{Z}}
\newcommand{\eps}{\varepsilon}
\newcommand{\bigO}{{\mathcal O}}

\newcommand\mapsfrom{\mathrel{\reflectbox{\ensuremath{\mapsto}}}}
\newcommand{\nbins}{n_{\text{bins}}}              % melody's parameters...
\newcommand{\bshared}{b^{\text{shared}}}
\newcommand{\Msub}{M_\text{sub}}

\newcommand{\psiper}{\psi^{\mathrm{per}}}   % note informative name

\definecolor{purple}{rgb}{0.5,0.0,1.0}
\def\compdata{data/compdata-1221} % define data dir (apart from Johannes data)

\pgfplotsset{filter discard warning=false}  % alex added to kill huge slow msgs

\definecolor{col-nu}{RGB}{255,0,0} % oranged red
\definecolor{col-sb}{RGB}{0,0,255} % dodgerblue
\definecolor{col-sb-2}{rgb}{0.13, 0.67, 0.8}
\definecolor{col-cufinufft}{RGB}{0,0,255} 
\definecolor{col-cufinufft-2}{rgb}{0.13, 0.67, 0.8}
\definecolor{col-finufft}{RGB}{255,30,0} 
\definecolor{col-cunfft}{RGB}{0,100,50}
\definecolor{col-gpunufft}{rgb}{0.44,0.16,0.39}
\pgfplotsset{
    discard if not/.style 2 args={
        x filter/.append code={
            \edef\tempa{\thisrow{#1}}
            \edef\tempb{#2}
            \ifx\tempa\tempb
            \else
                
            \fi
        }
    }
}
\pgfplotsset{
    discard if/.style 2 args={
        x filter/.append code={
            \edef\tempa{\thisrow{#1}}
            \edef\tempb{#2}
            \ifx\tempa\tempb
                
            \else

            \fi
        }
    }
}
\title{cuFINUFFT: a load-balanced GPU library for general-purpose nonuniform FFTs
}

\makeatletter
\newcommand{\linebreakand}{%
  \end{@IEEEauthorhalign}
  \hfill\mbox{}\par
  \mbox{}\hfill\begin{@IEEEauthorhalign}
}
\makeatother

\author{ %a
  \IEEEauthorblockN{Yu-hsuan Shih}
  \IEEEauthorblockA{\textit{Courant Institute of Mathematical Sciences}\\
    \textit{New York University}\\
    New York, NY, USA} % add email address
  \and
  \IEEEauthorblockN{Garrett Wright}
  \IEEEauthorblockA{\textit{PACM}\\ %Program in Applied and Computational Mathematics}\\
    \textit{Princeton University}\\
    Princeton, NJ, USA} % gbwright@princeton.edu
  \and
  \IEEEauthorblockN{Joakim And\'en}
  \IEEEauthorblockA{\textit{Department of Mathematics}\\
    \textit{KTH Royal Institute of Technology}\\
    Stockholm, Sweden} % janden@kth.se
  \linebreakand
  \IEEEauthorblockN{Johannes Blaschke}
  \IEEEauthorblockA{\textit{National Energy Research Scientific Computing Center}\\
    \textit{Lawrence Berkeley National Laboratory}\\
    Berkeley, CA, USA}
  \and
  \IEEEauthorblockN{Alex H. Barnett}
  \IEEEauthorblockA{\textit{Center for Computational Mathematics}\\
    \textit{Flatiron Institute}\\
    New York, NY, USA}
}  %a

\begin{document}
\maketitle

\begin{abstract}
  Nonuniform fast Fourier transforms dominate the computational cost in
  many applications including image reconstruction and signal processing.
  We thus
  present a general-purpose GPU-based CUDA library for type 1 (nonuniform to uniform)
  and type 2 (uniform to nonuniform) transforms in dimensions 2 and 3,
  in single or double precision.
 %
 % We optimize algorithm parameters to
 % achieve the user-requested accuracy as efficiently as possible.
  %
  It achieves high performance for a given user-requested accuracy,
  regardless of the distribution
  of nonuniform points, via
  cache-aware %nonuniform
  point reordering,
  and %, for type 1,
  load-balanced blocked spreading in shared memory.
  At low accuracies, this gives on-GPU throughputs around $10^9$ nonuniform points per second,
  % 2d & 3d, both types.
  %
%  We benchmark against three established codes over accuracies
%  from $10^{-2}$ to $10^{-12}$. 
  %
  and (even
  including
  %GPU memory allocation and
  host-device transfer)
  is typically 4--10$\times$     % instead of averages 6-8x.
  faster than the
  latest    % v2.0.2 includes Rob's better high-thread count scaling.
  parallel CPU code FINUFFT
  (at 28 threads).
  It is competitive with two established GPU codes,
  being up to 90$\times$ faster at high accuracy and/or type 1 clustered
  %being on average 90$\times$ faster at high accuracy and/or type 1 clustered
  point distributions.
  % *** 300x faster than the faster or slower of the two GPU codes? :)
  % 90x faster than the faster of the two GPU codes (sorry, I did a wrong calculation)
  Finally we demonstrate a 5--12$\times$ speedup versus CPU in an X-ray diffraction
    3D iterative reconstruction task at $10^{-12}$ accuracy, observing excellent multi-GPU weak scaling up to one rank per GPU.
% NERSC   ... but not only
  %single-node
\end{abstract}

% keywords can be removed

\section{Introduction}
%repeat NUFFT but points location change; repeat NUFFT with points location
%fixed. 

Nonuniform (or nonequispaced) fast Fourier transforms (NUFFTs)
are fast algorithms that generalize the FFT to the case of off-grid points.
%2D and 3D NUFFTs
They thus have a wealth of applications in engineering and scientific computing,
including image reconstruction from off-grid Fourier data
(e.g.\ MRI gridding \cite{fessler}, optical coherence tomography
\cite{octnufft},
cryo electron microscopy \cite{wang13cryo,cryo,Strelak2019,anden2018structural,
zhao2016fast},
radioastronomy \cite{arras20},
coherent diffraction X-ray imaging \cite{donatelli2017});
wave diffraction \cite{fresnaq};
partial differential equations \cite{gimbutasgrid,ludvig_nufft};
and long-range interactions in molecular \cite{salomon2013routine} and particle dynamics \cite{fiore2017}.
For reviews, see \cite{dutt,nufft,usingnfft,Barnett_2019}.

In 2D, the type 1 NUFFT \cite{dutt} (also known as the ``adjoint nonequispaced fast Fourier transform'' or ``adjoint NFFT'' \cite{usingnfft})
evaluates the uniform $N_1\times N_2$ grid of {\em Fourier series coefficients}
$f_{k_1,k_2}$
due to a set of point masses of arbitrary strengths $c_j$ and locations
$(x_j, y_j) \in [-\pi, \pi)^2$, $j=1,\dots,M$:
\begin{equation}
  f_{k_1, k_2} \defeq \sum_{j=1}^{M} c_j e^{-i (k_1x_j + k_2y_j)}
  ~, \quad (k_1, k_2) \in \mathcal{I}_{N_1, N_2},
	\label{eq:2d1}    % ugh, can you use shorter labels eg {2d1} ?
\end{equation}
where the 1D integer Fourier frequency grid is
%as with the FFT,
defined by
\be
\mathcal{I}_{N}\defeq \{k\in\ZZ: -N/2 \le k < N/2\}
~,
\label{I}
\ee
and we use the notation
$\mathcal{I}_{N_1,N_2} \defeq \mathcal{I}_{N_1}\times \mathcal{I}_{N_2}$
for a 2D grid of Fourier frequencies.

The type 2 NUFFT (or ``NFFT'' \cite{usingnfft})
is the adjoint of the type 1.
Given a grid of Fourier coefficients $f_{k_1,k_2}$
it evaluates the resulting Fourier series at arbitrary (generally nonuniform) targets $(x_j, y_j) \in [-\pi, \pi)^2$, to give
\begin{equation}
     c_j \defeq \sum_{(k_1, k_2)\in\mathcal{I}_{N_1, N_2}} f_{k_1,k_2} e^{i (k_1x_j
       +k_2y_j)}
     ~, \quad j = 1, \dots, M.\label{eq:2d2}
\end{equation}
In contrast to the FFT, type 1 is generally {\em not} the inverse of type 2:
inverting a NUFFT usually requires iterative solution of a
linear system \cite{fastsinc,usingnfft}.
Definitions \eqref{eq:2d1} and \eqref{eq:2d2} generalize to 1D and 3D
in the obvious fashion \cite{dutt}.

Naively the exponential sums in \eqref{eq:2d1} and \eqref{eq:2d2}
take $\bigO(NM)$ effort, where $N:=N_1\times\cdots\times N_d$ is the total number
of Fourier modes, and $d$ the dimension.
The NUFFT uses {\em fast algorithms} \cite{dutt,steidl98}
to approximate these sums to a user-prescribed
tolerance $\eps$, typically with effort of only
$\bigO(N\log N + M \log^d (1/\eps))$, i.e., quasi-linear in the data size.
Most algorithms internally set up a
``fine'' grid of size $n_1\times \dots \times n_d$, where each $n_i=\sigma N_i$, for a given upsampling factor $\sigma>1$.
Then the type 1 transform has three steps:
\bi
\item[i)]
  {\em spreading} (convolution) of each weighted nonuniform point
by a localized kernel, writing into the fine grid,
\item[ii)] performing a $d$-dimensional FFT of this fine grid, then
\item[iii)]
  selecting the central $N$ output modes from this fine grid,
  after pointwise division (deconvolution)
  by the kernel Fourier coefficients.
  %  correction for the filtering caused by the kernel,
\ei
Type 2 simply reverses (transposes) these steps, with i) becoming
kernel-weighted
{\em interpolation} from the fine grid to the nonuniform target points.
See Sec.~\ref{sec:algorithm} for details.

The NUFFT is often the rate-limiting step in applications, especially for
iterative reconstruction \cite{fessler},
motivating the need for high throughput.
Spreading and interpolation are often the dominant steps of NUFFTs, due
to scattered memory writes and reads of kernel-sized blocks.
Since it demands high memory bandwidth, yet is data parallel, it is a
task well suited for acceleration by a general-purpose GPU \cite{cheng2014}.

This potential of GPUs to accelerate the NUFFT has of course been noted,
and to some extent exploited, in prior implementations
\cite{cunfft,lin2018,gpunufft,gai2013}.
However, the present work shows that it is possible to increase the
efficiency significantly beyond that of prior codes, in the same hardware,
via algorithmic innovations.
With few exceptions \cite{cunfft},
most prior GPU NUFFT implementations are packaged in a manner
specific to a single science application
(e.g. MRI \cite{gai2013,gpunufft,lin2018}, OCT \cite{octnufft},
MD \cite{salomon2013routine}, or cryo-EM
\cite{Strelak2019}), have unknown or limited accuracy, and
lack mathematical documentation and testing, rendering them almost inaccessible
to the general user.
This motivates the need for an efficient, tested, general-purpose GPU code.

% CCCCCCCCCCCCCCCCCCCCCCCCCCCCCCCCCCCCCCCCCCCCCCCCCCCCCCCCCCCCCCCCCCCCCCCCCCCCC
\subsection{Contributions of this work}
\label{s:contrib}

%\begin{itemize}
%\item For \textit{spreading} in type 1 NUFFT, we propose two load-balanced methods,
%\textbf{GM-sort} and \textbf{SM}. Using the definition in
%\cite{Gregerson2008}, 
%\textbf{GM-sort} is an input (nonuniform point) driven approach, as those
%implementing in \cite{Kunis2012,Nestler2016,fiore2017} and \textbf{SM}
%is an output (grid) driven approach, as
%\cite{Gregerson2008,gai2013,Strelak2019,smith2019}). 
%Different from existing implementations, for both the methods, we bin-sort 
%the nonuniform points to regularize the memory access pattern. On top of that,
%in \textbf{SM}, we exploit GPU shared memrory and does local spreadings to 
%avoid slow atomic operations. 
%\item For \textit{interpolation} in type 2 NUFFT, we use method
%\textbf{GM-sort}, similar to the one in spreading. Bin-sorting nonuniform
%ponits introduces extra sorting cost while speeds up the actual interpolation
%kernel. It hence achieves best peroformance when there are multiple strength 
%vectors for the same nonuniform points. 
%\item The library use the new ``exponential of semicircle'' (ES) kernel
%that is proven to accelerate CPU implementation \cite{Barnett_2019}.
%\end{itemize}
We present {cuFINUFFT}, a
general-purpose GPU-based CUDA NUFFT library.
It exploits a recently-developed kernel with optimally small width
for a full range of user-chosen tolerances ($10^{-1}$ to $10^{-12}$),
yet more efficient to evaluate than prior ones \cite{Barnett_2019,keranal}.
Its throughput is high---and largely insensitive to the
point distribution.

One main contribution is to accelerate spreading in the type 1 NUFFT.
Broadly speaking there have been two styles of parallelization in prior work:
``input driven'' \cite{Gregerson2008} (or scatter \cite{Strelak2019}),
which assigns one
%\cite{cunfft,yang2018} or more \cite{fiore2017}
thread to each nonuniform point,
and
``output driven'' \cite{usingnfft,gai2013,Strelak2019,smith2019}
% *** can you check these refs use output-driven in GPU setting
% M: checked.
(gather), which assigns each thread
a distinct portion of the fine output grid to spread to.
The input driven scheme, accumulating
to GPU global memory, has been used in many prior GPU codes \cite{cunfft,yang2018,fiore2017};
we will refer to our implementation of this baseline method as \textbf{GM} (global memory).
While load-balanced, the memory access is arbitrary and can suffer from atomic collisions between writes
(see Sec.~\ref{s:spreadperf}).
Yet, a naive output driven approach, although collision-free,
is poorly load-balanced for highly nonuniform
point distributions \cite[Rmk.~12]{Barnett_2019}.
To address these issues we propose two new spreading methods:
%Thus we propose two load-balanced spreading methods:
%both with hand-tuned bin size to achieve high performance,
\begin{itemize}

\item \textbf{GM-sort} (global memory, sorted). This improves upon \textbf{GM} in that
%  An input driven scheme writing to global memory, but
  the work order of the nonuniform points is chosen by spatial
  sorting into bins (boxes) covering the fine grid.
  This regularizes the memory access pattern,
  enabling cache reuse.
  
\item \textbf{SM} (shared memory). This sets up spreading ``subproblems'' executed
  in faster GPU shared memory.
  %where atomic operations are faster.
  This is a hybrid scheme (see Fig.~\ref{fig:spreadoverview}):
  each subproblem has a local copy of the fine grid lying within the
  kernel half-width of one bin (output driven), yet is load-balanced
  by capping its subset of nonuniform points (input driven).
  %Once each subproblem is completed, its
  Local fine grids are added back into
  global memory using far fewer global atomic operations
  than \textbf{GM} methods, avoiding collisions.
  The result is 2--10$\times$ faster
  than \textbf{GM-sort} (depending on $d$ and clustering).
  %with throughputs that, as far as our comparisons can tell, are unmatched.
  
\end{itemize}
Bin sizes and shapes have been hand-tuned
for performance; this is crucial for \textbf{SM} to ensure optimal use of limited shared memory.

Turning to the interpolation task in type 2, we propose to use the adjoint version of \textbf{GM-sort}, where grid writes are replaced by reads.
We will also refer to this algorithm as \textbf{GM-sort}.
%(There is little need for \textbf{SM} version, due to the
%speed and lack of collision in global reads.)

For both tasks, while bin-sorting nonuniform points adds time,
it accelerates the execution of
spreading/interpolation.
Thus our library uses a ``plan, setup, execute, destroy'' interface that
allows efficient {\em reuse} of the same (sorted) nonuniform points with new strength
vectors (e.g.\ new $c_i$ in \eqref{eq:2d1}).
This use case is common, e.g.\ in iterative methods for NUFFT inversion.

We benchmark in detail the speedup of {cuFINUFFT} over existing NUFFT libraries,
for a range of accuracies, problem sizes, and point distributions.
For example, including GPU memory allocation and transfer
time, for low accuracy and quasi-uniform points,
{cuFINUFFT} is 
on average 8$\times$ faster than {FINUFFT} \cite{Barnett_2019} (28 threads),
5$\times$ faster than {CUNFFT}\cite{cunfft}
and 78$\times$ faster than {gpuNUFFT}\cite{gpunufft} for type 1 transforms.
For type 2,
{cuFINUFFT} is on average 6$\times$ faster than {FINUFFT}, 
5$\times$ faster than {gpuNUFFT},
and performs similarly to {CUNFFT} but with 2--5$\times$
faster ``execute'' times.
%(see Sec.  \ref{sec:bc} for definition of ``exec''). - note the above obviates this.

The library also enables multi-GPU parallelism,
essential for larger problems in HPC environments.
In Sec.~\ref{s:multi} we show this
in the setting of 3D single particle reconstruction
from coherent X-ray diffraction data, which demands thousands of 3D NUFFTs.
For NSERC and OLCF supercomputer nodes,
we demonstrate an order of magnitude
% *** fix when have multi-GPU!
speedup over the CPU version,
and excellent weak scaling with respect to the number of
GPU processes, up to one process per GPU.

The code and documentation for the library is available on GitHub\footnote{\url{https://github.com/flatironinstitute/cufinufft}} and installable as a PyPI package in Python through {\tt pip install cufinufft}.

% LLLLLLLLLLLLLLLLLLLLLLLLLLLLLLLLLLLLLLLLLLLLLLLLLLLLLLLLLLLLLLLLLLLLLLLL
\subsection{Limitations}
\label{s:limit}

Our library has a few limitations. (1) Both \textbf{GM-sort} and \textbf{SM} 
have some GPU memory overhead, due to sorting index arrays.
Yet, for a large 3D transform ($N_i=256,~i=1,2,3$, and
$M=1.3\times 10^8$), %512^3$,
this overhead is only around $20\%$.
% cut since we solve the problem!
%(2) The \textbf{GM-sort} method, while improving the global memory bandwidth,
%can slow down with clustered points due to write collisions
%(this motivated the \textbf{SM} method).
(2) The \textbf{SM} method, while providing a large acceleration for type 1,
is currently limited to single precision, due to the small
GPU shared memory per thread block (49 kB).
% could include double as a Future goal at end...
%
(3) We fixed the upsampling factor $\sigma=2$;
reducing this could reduce memory overhead and run times \cite{Barnett_2019}.
(4) Our library does not yet provide NUFFTs in 1D, nor of type 3 (nonuniform to nonuniform) \cite{nufft3}.
% *** any Johannes limitations?

% AAAAAAAAAAAAAAAAAAAAAAAAAAAAAAAAAAAAAAAAAAAAAAAAAAAAAAAAAAAAAAAAAAAAAAAAAA
\section{Algorithms}
\label{sec:algorithm}
We follow the standard three-step scheme presented in the previous section.
Our Fourier transform convention is
\begin{equation}
	\hat{\phi} (k) = \int_{-\infty}^{\infty} \!\!\!\phi(x)
	e^{-ikx}dx
        \,,\;
	\phi (x) = \frac{1}{2\pi}\int_{-\infty}^{\infty}\!\! \hat{\phi}(k)
	e^{ikx}dk.
\end{equation}
We fix the upsampling factor $\sigma = 2$,
and use the ``exponential of semicircle'' (ES) kernel from FINUFFT \cite{Barnett_2019,keranal},
\begin{equation}
    \phi_\beta(z)\defeq
    \begin{cases}
    e^{\beta(\sqrt{1-z^2}-1)}, &|z|\leq 1\\
    0, & \text{otherwise}.
    \end{cases}
	\label{es}
\end{equation}
%The kernel parameters are designed to
%achieve relative $l_2$ tolerance $\varepsilon$.
Given a user-requested tolerance $\varepsilon$,
the kernel width $w$ in fine grid points, and parameter $\beta$ in \eqref{es},
are set via
\begin{equation}
	w = \lceil \log_{10} 1/\varepsilon \rceil+1, \quad \beta = 2.30w.
\end{equation}
This typically gives relative $\ell_2$ errors close to $\varepsilon$ \cite{Barnett_2019}.
As in FINUFFT, for FFT efficiency, 
the fine grid size $n_i$ is set to be
the smallest integer of the form $2^{q_i}3^{p_i}5^{r_i}$,
greater than or equal to $\max\left(\sigma N_i, 2w\right)$, in each
%not less than $\sigma N_i$ nor $2w$
dimension $i=1,\dots,d$. %for $w$ being the kernel width.

\subsection{Type 1: nonuniform to uniform}
\label{algo:type1}
We use the same algorithm as FINUFFT to compute $\tilde{f}_{k_1,k_2}$, an
approximation to $f_{k_1,k_2}$ in \eqref{eq:2d1}.
We will write only the 2D case, the generalization to 3D being clear.

\paragraph{Step 1 (spreading)} For each index $(l_1,l_2)$
in the fine grid $0\le l_1<n_1$, $0\le l_2<n_2$, compute
\be
	b_{l_1, l_2} = \sum_{j=1}^M c_j \psiper(l_1h_1-x_j, l_2h_2-y_j),
\ee
where $h_i\defeq 2\pi/n_i$ is the fine grid spacing, and $\psiper(x,y)$
is the periodized tensor product of rescaled ES kernels
\begin{align}
	&\psi(x,y) :=
	\phi_\beta(x/\alpha_1)\phi_\beta(y/\alpha_2),\enskip \alpha_i
	=w\pi/n_i, \enskip i=1,2, \nonumber \\ 
	&\psiper(x,y) := \sum_{(m_1,m_2)\in
	\mathbb{Z}^2} \psi(x-2\pi m_1,y-2\pi m_2).
\end{align}
%Note that because $\psi$ is finitely supported, most of the terms in the sum are zero:
Note that
each nonuniform point $(x_j,y_j)$ only
affects a nearby square of $w^2$ fine grid points.
%$\psi(x-2\pi m_1,y-2\pi m_2)$
%that are nonzero.
\paragraph{Step 2} Use a plain 2D FFT to evaluate
\begin{multline}
	\hat{b}_{k_1,k_2} = \sum_{l_2 = 0}^{n_2-1}\sum_{l_1 = 0}^{n_1-1}
b_{l_1,l_2}e^{-2\pi i (l_1k_1/n_1 + l_2k_2/n_2)},\\ (k_1,k_2)\in
\mathcal{I}_{n_1,n_2}.
\end{multline}
\paragraph{Step 3 (correction)} Truncate the Fourier coefficients to
the central $N_1\times N_2$ frequencies, and scale them
to give the final outputs
\begin{equation}
	\tilde{f}_{k_1,k_2} = p_{k_1,k_2}\hat{b}_{k_1,k_2}, \quad
	(k_1,k_2)\in \mathcal{I}_{N_1,N_2}.
\end{equation}
Here, correction (deconvolution) factors $p_{k_1,k_2}$ are
precomputed from samples of the kernel Fourier transform, via
$$
p_{k_1,k_2} = h_1h_2\hat{\psi}(k_1,k_2)^{-1} =
(2/w)^2(\hat{\phi}_\beta(\alpha_1k_1)\hat{\phi}_\beta(\alpha_2k_2))^{-1}.
$$

\subsection{Type 2: uniform to nonuniform}
\label{algo:type2}

To compute $\tilde{c}_j$, an approximation to ${c}_j$ in \eqref{eq:2d2}, as in
FINUFFT, the
above steps for type 1 are reversed. 
The correction factors $p_{k_1,k_2}$ and the
periodized kernel $\psiper$ remain as above.
\paragraph{Step 1 (correction)} Pre-correct then zero-pad the coefficients
$f_{k_1,k_2}$ to the fine grid, i.e., for all indices $(l_1,l_2)$,
\be
    \hat{b}_{l_1, l_2} =
    \begin{cases}
      p_{k_1,k_2}f_{k_1,k_2}, & (k_1,k_2)\in \mathcal{I}_{N_1,N_2}\\
      0, & (k_1,k_2)\in \mathcal{I}_{n_1,n_2}\backslash \mathcal{I}_{N_1,N_2}
    \end{cases}\\
\ee
\paragraph{Step 2} Use a plain inverse 2D FFT to evaluate
\begin{multline}
	b_{l_1,l_2} = \sum_{(k_1,k_2)\in \mathcal{I}_{n_1,n_2}}
	\hat{b}_{k_1,k_2}e^{2\pi i (l_1k_1/n_1 + l_2k_2/n_2)},\\ 
	l_i =0, \dots, n_i-1, \enskip i=1,2.
\end{multline}
\paragraph{Step 3 (interpolation)} Compute a weighted sum of the $w^2$ grid values near each target nonuniform point $(x_j,y_j)$,
$$
\tilde{c}_j = \sum_{l_1 = 0}^{n_1-1}\sum_{l_2 = 0}^{n_2-1}
	b_{l_1,l_2}\psiper(l_1h_1-x_j,l_2h_2-y_j),\; j=1,\cdots,M.
$$

% GGGGGGGGGGGGGGGGGGGGGGGGGGGGGGGGGGGGGGGGGGGGGGGGGGGGGGGGGGGGGGGGGGGGGGGG
\section{GPU implementation}

This section shows how the above three-step algorithms are implemented on the GPU using
the CUDA API.
For the FFT in both types, we use NVIDIA's {cuFFT}
library. For the correction steps, since the task
is embarrassingly parallel, we simply launch one thread for each
of the $N_1\times N_2$ Fourier modes. The factors $p_{k_1,k_2}$ are precomputed
once in the planning stage.
The major work lies in the spreading (type 1) and interpolation (type 2),
to which we now turn.

\begin{figure*}[ht] %fffffffffffffffffffffffffffffffffffffffffffffffffffffffffff
  \centering
  \includegraphics[width=\textwidth]{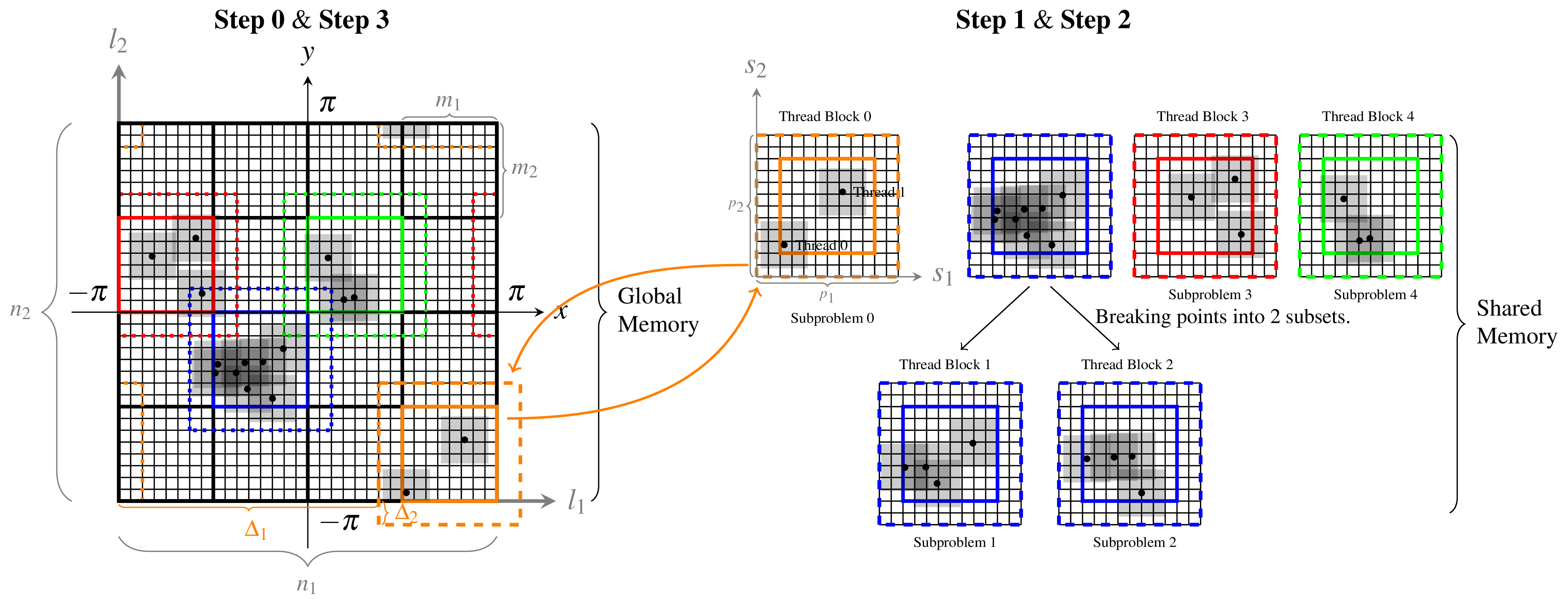}
	\caption{\textbf{SM} spreading scheme.
          The shaded
          gray squares (each $w \times w$ fine grid points) show the support of the spreading kernel, centered on each nonuniform point (black dots).
          Colors indicate the different nonempty bins.
	\textbf{Step 0:} Divide the $n_1\times n_2$
	fine grid into bins of size $m_1\times m_2$. 
	\textbf{Step 1:} Assign subproblems by bin-sorting non-uniform points, 
	and, if needed, further splitting so that there are at most $\Msub$ points per subproblem.
	\textbf{Step 2:} Spread the points inside each subproblem into
	a padded bin copy in faster shared memory (size $p_1\times p_2$).
	\textbf{Step 3:} Atomic add each padded bin result back into global memory.
	}
  \label{fig:spreadoverview}
\end{figure*}

% SSSSSSSSSSSSSSSSSSSSSSSSSSSSSSSSSSSSSSSSSSSSSSSSSSSSSSSSSSSSSSSSSSSSSSSS
\subsection{Spreading}
\label{sec:spreading}
%\msnote{\textbf{GM, GM-sort, SM.}
%\begin{itemize}
%	\item (Motivation) Downsides of the baseline method \textbf{GM}: 
%		(1) Random write to fine grid data in global memory.
%		(2) Slow atomic operations.  
%	\item \textbf{GM-sort}: Coalesce memory writes by bin-sorting the
%		nonuniform points.  
%	\item \textbf{SM}: bin-sort the nonuniform points and does local spreading
%		to shared memory to reduce global atomic operations. Load-balance is
%		achieved by dividing nonuniform points inside an output bin into
%		subsets. Work of spreading each set of points is called a subproblem.
%\end{itemize}
%}

%As mentioned in Sec.~\ref{s:contrib},
Recall that we use \textbf{GM} to denote
a baseline input driven spreading implementation in global memory
(as used in CUNFFT \cite{cunfft}).
This launches one thread per nonuniform point,
i.e.\ $M$ in total, in their user-supplied order.
The thread given nonuniform point $j$ spreads it to the fine grid $b$ array:
$$
b_{l_1, l_2} \;\mapsfrom\;
b_{l_1, l_2} + c_j \psiper(l_1h_1-x_j,l_2h_2-y_j),
\quad \forall (l_1,l_2)~.
$$
This task comprises (a) reading $x_j$, $y_j$, $c_j$ from GPU global memory,
(b) $2w$ evaluations of the kernel function $\psi$, and (c) $w^2$ atomic adds to the $b$ array residing in GPU global memory.
%$b_{l_1,l_2}, l_i = 0,\dots,n_i-1, i=1,2$.
%Note that because
%threads might write into the same location of array $b$, the additions have to
%be done atomically and that again because $\psi$ has finite support, only $w^2$
%entries of $b$ array are modified. 
One downside of this approach is that nonuniform points assigned to threads
within a thread block and hence within a warp can reside far from each other on
the grid, which results in uncoalesced memory loads.
(Note that assigning one {\em thread block} per nonuniform point
may alleviate this issue \cite{fiore2017}.)
% *** Melody check. M:checked.
Another downside is that global atomic operations
for clustered points can essentially serialize the method.
% *** and this issue GM-sort does not fix, right? M: No, GM-sort doesn't fix this.

\textbf{GM-sort} is a scheme to address the issue of uncoalesced access.
We partition the $n_1\times n_2$ fine grid
into rectangular ``bins'' $R_i$, $i=1, \dots, \nbins$,
each of integer sizes $m_1\times m_2$ if
possible, otherwise smaller. (A typical choice is $m_1=m_2=32$.)
Thus $\nbins = \lceil\frac{n_1}{m_1}\rceil\times \lceil \frac{n_2}{m_2}\rceil$.
Bins are ordered in a Cartesian grid with the $x$ axis fast and  $y$ slow,
%(and in the 3D case, then $z$ slowest).
which echoes on a larger scale the ordering of the fine grid itself.
We will say that nonuniform point $j$ is ``inside'' bin $R_i$
if the point's rounded integer fine grid coordinates
$$
l_1 = \lfloor (x_j+\pi)/h_1 \rfloor ~, \qquad
l_2 = \lfloor (y_j+\pi)/h_2 \rfloor ~,
$$
lie in $R_i$.
Suppose that there are $M_i$ nonuniform points inside bin $R_i$ for
$i=1,\dots, \nbins$.
We set up a permutation $t$ (a bijection from $\{1,\dots,M\}$ to itself),
such that the nonuniform points with indices $t(1),t(2),\dots,t(M_1)$
are precisely those lying in bin $R_1$,
then those with indices $t(M_1+1),t(M_1+2),\dots,t(M_1+M_2)$ are precisely
those in bin $R_2$, etc.
This is done in practice by first recording the bin index of
each point, reading out this list in bin ordering, then inverting
this permutation to give $t$.
% *** say how bin-sort parallized on GPU ?
Nonuniform points are then assigned to threads
in the permuted index order $t(1),\dots,t(M)$.
%We permute the nonuniform points using an index $t_j$
%and assign the nonuniform points to threads in the sorted order. 
%assigned them to threads $t_j,\enskip j=1,\dots,M$
%in the sorted order 
%$(x_{t_j},y_{t_j})$, $j=1,\dots,M$  
%\begin{multline} 
%	b_{l_1, l_2} = b_{l_1, l_2} + c_{t_j} \psiper(l_1h_1-x_{t_j},l_2h_2-y_{t_j}),\\
%	l_i = 0, \dots, n_i-1,\enskip i=1,2.
%\end{multline} 
This means that the threads within a warp now access parts of the $b$
fine grid array that, most of the time, are approximately adjacent.
The GPU therefore has a better chance of
coalescing these accesses into fewer global memory transactions.

\textbf{SM} is a hybrid scheme which exploits GPU shared memory
to further address the issue of slow global atomic operations.
It partitions the fine grid into bins $R_i$ as above,
then has three remaining steps, as follows (see Fig.~\ref{fig:spreadoverview}).

\textit{Step 1: Assign subproblems using bin-sorting and blocking
  of nonuniform points}.
The nonuniform point index list $1,\dots,M$ is broken into the union of
disjoint subsets $S_1,S_2,\dots$, each of which we call a ``subproblem''.
This is done as follows. For bin $R_1$, if the number of points $M_1$ is larger than $\Msub$, a parameter controlling the maximum subproblem size, then it is further broken into
subsets (subproblems)
of size at most $\Msub$. These one or more subproblems are all
associated to the bin $R_1$, in which their points lie.
The same is repeated for the remainder of the bins $R_i$.
Thread blocks are then launched, one per subproblem.
Note that the cap on subproblem size is a (blocked)
form of input driven load-balancing: if many points lie in a bin,
their spreading tasks are well parallelized.  % **** bit clumsy

\textit{Step 2: Spread nonuniform points inside each subproblem to shared
memory.} By the above partition, within a subproblem (a thread block), the
nonuniform points can only affect the fine grid array $b$
within a {\em padded bin} of dimensions $p_1\times p_2$, where
\begin{equation}
	p_i = \left(m_i + 2 \lceil w/2\rceil\right),\quad i=1,2.
\end{equation}
For the $k$th subproblem $S_k$, we
accumulate its spreading result on a shared memory
copy $\bshared$ of its padded bin, local to its thread block,
\begin{multline}
	\bshared_{s_1, s_2} \;\;\mapsfrom\;\; \bshared_{s_1, s_2} \; +\\
	\sum_{j \in S_k} c_j \psiper((s_1+\Delta_1)h_1-x_j,(s_2+\Delta_2)h_2-y_j),\\
	s_i = 0,\dots,p_i-1,\quad i=1,2,
\end{multline}
where $(s_1,s_2)$ are local indices within the padded bin copy,
and $(\Delta_1,\Delta_2)$ its offset within the fine grid
(see Fig.~\ref{fig:spreadoverview}).

\textit{Step 3: Atomic add the results back to global memory.} Once the
spreading subproblem result is accumulated in the shared memory padded bin, we atomically add it back to the corresponding region of global memory array $b$,
\be
b_{l_1(s_1),l_2(s_2)} \;\mapsfrom\;
b_{l_1(s_1),l_2(s_2)}
  + \bshared_{s_1,s_2}, \enskip \forall (s_1,s_2)
%	s_i = 0,\dots,p_i-1,\enskip i=1,2,
    \label{eq:addtoglobal}
\ee
where $l_i(s_i) := (s_i+\Delta_i)\, \mbox{mod}\, n_i$, $i=1,2$, maps
each coordinate in the padded bin back to the fine grid,
with periodic wrapping (see Fig.~\ref{fig:spreadoverview}).
This completes the spreading process. 
When there are many nonuniform points per bin, this incurs
many fewer global atomic writes than \textbf{GM-sort}.

We generalize both the implementations \textbf{GM-sort} and \textbf{SM} to 3D 
by using cuboid bins of maximum dimension $m_1\times m_2\times m_3$.

\begin{rem}
  In both methods \textbf{GM-sort} and \textbf{SM}, the performance is
  sensitive to the bin sizes. By hand-tuning (in powers of two), we have found $32\times 32$ in 2D and $16\times16\times2$ in 3D to be optimal.
This takes account of the area (or volume) ratio of bin to padded bin,
the ordering of fine grid data, and the maximum size of GPU shared memory per thread block (49 kB), and are based on speed tests on an NVIDIA Tesla V100.
We similarly set $\Msub=1024$, although we believe that its optimal
value is problem-dependent.
\end{rem}

\begin{rem}
  We implemented \textbf{SM} in both dimensions and precisions,
  apart from 3D double precision where it is no longer advantageous.
  Here the shared memory constraint
  $16(m_1+w)(m_2+w)(m_3+w) \le 49000$,
  for widths $w>8$ needed when $\varepsilon<10^{-7}$,
  forces the bin volume to be tiny compared to the padded bin volume,
  dramatically increasing the number of global operations.
  We test only \textbf{GM-sort} in this case.
  % MATLAB/Octave calc of this: it can be done, but efficiency is low:
  %> m=[8,8,2]; w=8; eff=prod(m)/prod(m+w), SM=16*prod(m+w)
  %eff =  0.050000
  %SM =  40960
\end{rem}

% IIIIIIIIIIIIIIIIIIIIIIIIIIIIIIIIIIIIIIIIIIIIIIIIIIIIIIIIIIIIIIIIIIIIIIIIII
\subsection{Interpolation}
\label{sec:interpolation}
%\msnote{\textbf{GM-sort}.}
%\msnote{Same idea as \textbf{GM-sort} in spreading with read/write memory 
%operations reversed. Since there is no confliction of threads reading same location
%of memory, the benefit of applying \textbf{SM} by loading first data into
%the shared memory is limited.} 
For interpolation, we use the same idea as \textbf{GM-sort} for spreading,
with read and write memory operations reversed.
Threads are assigned to nonuniform points in the permuted order
$t(1),\dots,t(M)$ coming from bin-sorting.
Thus the $j$th thread performs the task
$$
	\tilde{c}_{t(j)} = 
	\sum_{l_1 = 0}^{n_1-1}\sum_{l_2 = 0}^{n_2-1}
	%\sum_{\left\{(l_1,l_2)\left| \substack{ |l_1h_1-x_{t_i}| \leq \alpha_1 \\
	%							                    |l_2h_2-y_{t_i}| \leq \alpha_2}\right.\right\}}
	b_{l_1,l_2}\psiper(l_1h_1-x_{t(j)},l_2h_2-y_{t(j)})~.
$$
We refer to the method by the same name, \textbf{GM-sort}, and use 
\textbf{GM} to refer to 
the unsorted version. The reason to bin-sort the nonuniform points is
to coalesce the reads from the fine grid.
Since there is no conflict between threads reading the same
location in memory, this is fast;
the benefit of applying an idea like \textbf{SM}
to interpolation would be limited.

% PPPPPPPPPPPPPPPPPPPPPPPPPPPPPPPPPPPPPPPPPPPPPPPPPPPPPPPPPPPPPPPPPPPPPPPPPP
\section{Performance Tests}
\label{sec:performance}

In this section, we report the performance of our GPU library,
{cuFINUFFT}. We first show how the proposed spreading methods
\textbf{GM-sort} and \textbf{SM}
perform against \textbf{GM}.
Then, we compare the performance of the interpolation
step with (\textbf{GM-sort}) and without (\textbf{GM}) bin-sorting the nonuniform points.
Finally, we show how the whole pipeline performance
(spreading/interpolation, FFT, correction) of cuFINUFFT compares with
the fastest known multithreaded CPU library {FINUFFT} \cite{Barnett_2019},
and two established GPU libraries {CUNFFT} \cite{cunfft} and {gpuNUFFT} \cite{Strelak2019}.  

All GPU timings are for a NVIDIA Tesla V100 (released in 2017),
with memory bandwidth 900 GB/s.
%(See Sec.~\ref{sec:bc} for the comparison CPU used.)
We compile all codes with GCC v7.4.0 and NVCC
v10.0.130.  Unless specified, single precision and $\Msub=1024$ are 
used in all the tests.

\textbf{\textit{Tasks.}} We present results for the following two
nonuniform point distributions, which are extreme cases:
\begin{itemize}
	\item ``rand'': nonuniform points are independent and identically distributed uniform random variables 
		across the entire periodic domain box $[-\pi, \pi)^d$, $d=2,3$.
	\item ``cluster'': points are iid random in the small box $[0, 8h_1]\times\dots\times [0, 8h_d]$, recalling that $h_i$ are the fine grid spacings.
%		Note that this distribution only valids with $n_i\geq 16, i=1,2,3$. 
	%\item ``disc quad'':  a polar grid over the disc of radius $\pi$, using
	%	roughly $\sqrt{M}$ radial Gauss--Legendre nodes and $\sqrt{M}$
	%	equi-spaced angular nodes.
	%\item ``3D sph quad'': a spherical grid in the ball of radius $\pi$, using
	%	$\sqrt{M}/2$ radial Gauss--Legendre nodes and a $\sqrt{M}\times
	%	2\sqrt{M}$ tensor-product grid on each sphere.
\end{itemize}
%``rand'' and ``cluster'' are two extreme cases of nonuniform points'
%distributions. 
%and ``disc quad'' is a distribution that lies somewhere in between. 
We restrict to square/cube problems, i.e.\ $N_1=N_2(=N_3)$, being
the most common application problem.
We define problem {\em density} $\rho$ to be the ratio of number of nonuniform
points to number of upsampled grid points, i.e.,
\begin{equation}
	\label{eq:den}
	\rho \; := \; \frac{M}{\prod_{i=1}^d n_i} \;=\;  \frac{M}{\sigma^d \prod_{i=1}^d N_i}~.
\end{equation}
We report tests with $\rho$ of order 1, since
a) this is common in applications, and b) in the uncommon
case $\rho\ll 1$ one ends up essentially comparing plain FFT speeds. %dominate.
In fact we have tested $\rho=0.1$ and $\rho=10$, as well as less extreme
nonuniform point distributions; the conclusions are rather similar.

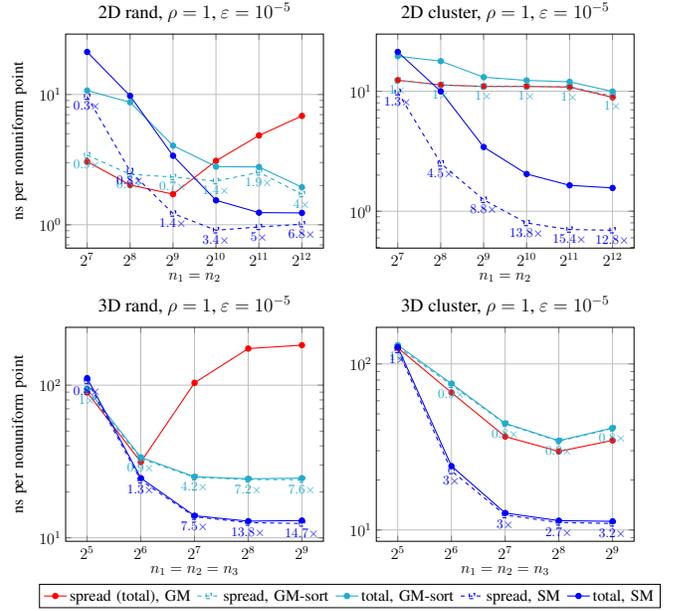
\begin{figure}[t]  % ffffffffffffffffffffffffffffffffffffffffffffffffffffff
  \centering
  % 4 plots here...

\begin{tikzpicture}[scale=0.5, transform shape]
	\begin{groupplot}[group style={{group size=2 by 2},
		vertical sep=6em,
		horizontal sep=4em},
		log basis x={2},
		grid,
		legend pos=north east,
		legend cell align=left,
		ymode=log,
		xmode=log,
		scale ticks above exponent={2},
		scaled y ticks=true,
		every node near coord/.append style={
			/pgf/number format/fixed,
			/pgf/number format/precision=1,}
		]
		%% 2D rand
		\nextgroupplot[title={2D rand, $\rho=1$, $\varepsilon=10^{-5}$},
		ylabel=ns per nonuniform point,
		xlabel={$n_1=n_2$},
		legend to name={spreadlegend},
		legend style={legend columns=-1},
		legend style={/tikz/every even column/.append style={column sep=0.1cm}}]
		\addplot +[discard if not={method}{0},
		discard if not={density}{1},
		discard if ={nf1}{32},
		discard if ={nf1}{64},
		col-nu,thick, 
		mark=*, mark options={fill=col-nu}]
		table [x=nf1,y
		expr={1e9*1e-3*(\thisrow{spread})/\thisrow{M}}]
		{data/spread2d_density_uniform_withspeedup.dat};
		\addlegendentry{spread (total), {GM}}
		\addplot +[discard if not={method}{1},
		discard if not={density}{1},
		discard if ={nf1}{32},
		discard if ={nf1}{64},
		col-sb-2,dashed,thick, 
		mark=square, mark options={fill=col-sb-2},
		point meta=explicit,
		nodes near coords=\pgfmathprintnumber{\pgfplotspointmeta}$\times$,
		nodes near coords align={below},
		]
		table [meta=speedup,x=nf1,y
		expr={1e9*1e-3*(\thisrow{spread})/\thisrow{M}}]{data/spread2d_density_uniform_withspeedup.dat};
		\addlegendentry{spread, {GM-sort}}
		\addplot +[discard if not={method}{1},
		discard if not={density}{1},
		discard if ={nf1}{32},
		discard if ={nf1}{64},
		col-sb-2,thick, 
		mark=*, mark options={fill=col-sb-2}]
		table [x=nf1,y
		expr={1e9*1e-3*(\thisrow{spreadprop}+\thisrow{spread})/\thisrow{M}}]{data/spread2d_density_uniform_withspeedup.dat};
		\addlegendentry{total, {GM-sort}}
		\addplot +[discard if not={method}{2},
		discard if not={density}{1},
		discard if ={nf1}{32},
		discard if ={nf1}{64},
		col-sb, dashed,
		thick,
		mark=square,mark options={fill=col-sb},
		point meta=explicit,
		nodes near coords=\pgfmathprintnumber{\pgfplotspointmeta}$\times$,
		nodes near coords align={below},
		]
		table [meta=speedup,x=nf1,y
		expr={1e9*1e-3*(\thisrow{spread})/\thisrow{M})}]{data/spread2d_density_uniform_withspeedup.dat};
		\addlegendentry{spread, {SM}}
		\addplot +[discard if not={method}{2},
		discard if not={density}{1},
		discard if ={nf1}{32},
		discard if ={nf1}{64},
		thick,
		col-sb,
		mark=*,mark options={fill=col-sb}]
		table [x=nf1,y
		expr={1e9*1e-3*(\thisrow{spreadprop}+\thisrow{spread})/\thisrow{M}}]{data/spread2d_density_uniform_withspeedup.dat};
		\addlegendentry{total, {SM}}

		%% 2D cluster
		\nextgroupplot[title={2D cluster, $\rho=1$, $\varepsilon=10^{-5}$},
		xlabel={$n_1=n_2$}]
		\addplot +[discard if not={method}{0},
		discard if not={density}{1},
		discard if ={nf1}{32},
		discard if ={nf1}{64},
		col-nu,thick, 
		mark=*, mark options={fill=col-nu}]
		table [x=nf1,y expr={1e9*1e-3*(\thisrow{spread})/\thisrow{M}}]{data/spread2d_density_cluster_withspeedup.dat};
		\addplot +[discard if not={method}{1},
		discard if not={density}{1},
		discard if ={nf1}{32},
		discard if ={nf1}{64},
		col-sb-2,dashed,thick, 
		point meta=explicit,
		nodes near coords=\pgfmathprintnumber{\pgfplotspointmeta}$\times$,
		nodes near coords align={below},
		mark=square, mark options={fill=col-sb-2}]
		table [meta=speedup,x=nf1,y
		expr={1e9*1e-3*(\thisrow{spread})/\thisrow{M}}]{data/spread2d_density_cluster_withspeedup.dat};
		\addplot +[discard if not={method}{1},
		discard if not={density}{1},
		discard if ={nf1}{32},
		discard if ={nf1}{64},
		col-sb-2,thick, 
		mark=*, mark options={fill=col-sb-2}
		]
		table [x=nf1,y
		expr={1e9*1e-3*(\thisrow{spreadprop}+\thisrow{spread})/\thisrow{M}}]{data/spread2d_density_cluster_withspeedup.dat};
		\addplot +[discard if not={method}{2},
		discard if not={density}{1},
		discard if ={nf1}{32},
		discard if ={nf1}{64},
		col-sb, dashed,
		mark=square,mark options={fill=col-sb},
		thick,
		point meta=explicit,
		nodes near coords=\pgfmathprintnumber{\pgfplotspointmeta}$\times$,
		nodes near coords align={below},]
		table [meta=speedup,x=nf1,y expr={1e9*1e-3*(\thisrow{spread})/\thisrow{M})}]{data/spread2d_density_cluster_withspeedup.dat};
		\addplot +[discard if not={method}{2},
		discard if not={density}{1},
		discard if ={nf1}{32},
		discard if ={nf1}{64},
		thick,
		col-sb,
		mark=*,mark options={fill=col-sb}]
		table [x=nf1,y expr={1e9*1e-3*(\thisrow{spreadprop}+\thisrow{spread})/\thisrow{M}}]{data/spread2d_density_cluster_withspeedup.dat};

		%% 3D rand
		\nextgroupplot[title={3D rand, $\rho=1$, $\varepsilon=10^{-5}$},
		ylabel=ns per nonuniform point,
		xlabel={$n_1=n_2=n_3$}]
		\addplot +[discard if not={method}{0},
		discard if not={density}{1},
		col-nu, solid, thick,
		mark=*, mark options={fill=col-nu}]
		table [x=nf1,y expr={1e9*1e-3*(\thisrow{spread})/\thisrow{M}}]{data/spread3d_density_uniform_withspeedup.dat};
		\addplot +[discard if not={method}{1},
		discard if not={density}{1},
		col-sb-2,dashed,thick, 
		mark=square, mark options={fill=col-sb-2},
		point meta=explicit,
		nodes near coords=\pgfmathprintnumber{\pgfplotspointmeta}$\times$,
		nodes near coords align={below},
		]
		table [meta=speedup,x=nf1,y
		expr={1e9*1e-3*(\thisrow{spread})/\thisrow{M}}]{data/spread3d_density_uniform_withspeedup.dat};
		\addplot +[discard if not={method}{1},
		discard if not={density}{1},
		col-sb-2,thick, 
		mark=*, mark options={fill=col-sb-2},
		]
		table [x=nf1,y
		expr={1e9*1e-3*(\thisrow{spreadprop}+\thisrow{spread})/\thisrow{M}}]{data/spread3d_density_uniform_withspeedup.dat};
		\addplot +[discard if not={method}{2},
		discard if not={density}{1},
		col-sb, dashed, thick,
		mark=square,mark options={fill=col-sb},
		point meta=explicit,
		nodes near coords=\pgfmathprintnumber{\pgfplotspointmeta}$\times$,
		nodes near coords align={below},]
		table [meta=speedup,x=nf1,y expr={1e9*1e-3*(\thisrow{spread})/\thisrow{M})}]{data/spread3d_density_uniform_withspeedup.dat};
		\addplot +[discard if not={method}{2},
		discard if not={density}{1},
		col-sb, thick,
		mark=*,mark options={fill=col-sb}]
		table [x=nf1,y expr={1e9*1e-3*(\thisrow{spreadprop}+\thisrow{spread})/\thisrow{M}}]{data/spread3d_density_uniform_withspeedup.dat};

		%% 3D cluster
		\nextgroupplot[title={3D cluster, $\rho=1$, $\varepsilon=10^{-5}$},
		xlabel={$n_1=n_2=n_3$},
		%ymax=1e-7*1e9,
		%ymin=1e-8*1e9
		]
		\addplot +[discard if not={method}{0},
		discard if not={density}{1},
		col-nu, solid, thick,
		mark=*, mark options={fill=col-nu}]
		table [x=nf1,y expr={1e9*1e-3*(\thisrow{spread})/\thisrow{M}}]{data/spread3d_density_cluster_withspeedup.dat};
		\addplot +[discard if not={method}{1},
		discard if not={density}{1},
		col-sb-2,dashed,thick, 
		mark=square, mark options={fill=col-sb-2},
		point meta=explicit,
		nodes near coords=\pgfmathprintnumber{\pgfplotspointmeta}$\times$,
		nodes near coords align={below},]	
		table [meta=speedup,x=nf1,y
		expr={1e9*1e-3*(\thisrow{spread})/\thisrow{M}}]{data/spread3d_density_cluster_withspeedup.dat};
		\addplot +[discard if not={method}{1},
		discard if not={density}{1},
		col-sb-2,thick, 
		mark=*, mark options={fill=col-sb-2}]
		table [x=nf1,y
		expr={1e9*1e-3*(\thisrow{spreadprop}+\thisrow{spread})/\thisrow{M}}]{data/spread3d_density_cluster_withspeedup.dat};
		\addplot +[discard if not={method}{2},
		discard if not={density}{1},
		col-sb, dashed, thick,
		mark=square,mark options={fill=col-sb},
		point meta=explicit,
		nodes near coords=\pgfmathprintnumber{\pgfplotspointmeta}$\times$,
		nodes near coords align={below},]
		table [meta=speedup,x=nf1,y expr={1e9*1e-3*(\thisrow{spread})/\thisrow{M})}]{data/spread3d_density_cluster_withspeedup.dat};
		\addplot +[discard if not={method}{2},
		discard if not={density}{1},
		col-sb, thick,
		mark=*,mark options={fill=col-sb}]
		table [x=nf1,y expr={1e9*1e-3*(\thisrow{spreadprop}+\thisrow{spread})/\thisrow{M}}]{data/spread3d_density_cluster_withspeedup.dat};
	\end{groupplot}
\path (group c1r2.west|-current bounding box.south) -- node[below]{\ref{spreadlegend}} (group c2r2.east|-current bounding box.south);
\end{tikzpicture}
  \caption{Spreading method comparisons. % at $\varepsilon=10^{-5}$.
    Execution time per nonuniform
	point (smaller is better) is shown, for various fine grid sizes,
	and distributions ``rand'' and ``cluster'', in 2D and 3D.
        For \textbf{GM-sort} (cyan) and \textbf{SM} (dark blue),
        the ``total'' time (solid lines) includes the precomputation
        (bin-sorting or subproblem setup),
        whereas the ``spread'' time (dotted lines)
        excludes this precomputation.
        The baseline \textbf{GM} (red) needs no precomputation.
        Annotations are the speedups over \textbf{GM}.}
  \label{spreading2D}
\end{figure}

% SSSSSSSSSSSSSSSSSSSSSSSSSSSSSSSSSSSSSSSSSSSSSSSSSSSSSSSSSSSSSSSSSSSSSSSSS
\subsection{Spreading performance}
\label{s:spreadperf}
%\msnote{
%\begin{itemize}
%	\item For large fine grid, bin-sorting nonuniform points becomes
%		crucial because if there is no controls of blocks of fine grid threads
%		to write to, those blocks can be far away from each other and it is not
%		possible for GPU to coalesce the memory accesses. 
%	\item On the other hand, for small fine grid, finer control is needed to
%		improve the usage of GPU global memory bandwidth. We see no benefit
%		of simply bin-sorting the points. 
%		%and even worse that since now threads
%		%are processing closer nonuniform points, more writing conflictions 
%		%can happen. 
%	\item For cluster distribution, nonuniform points all reside in a small 
%		region which allows us to focus on examining the issue of slow global
%		atomic operations. It is clear that using shared memory and doing local
%		spreadings first in \textbf{SM} improves the situation.
%\end{itemize}
%}

For high-accuracy single precision ($\varepsilon=10^{-5}$, i.e.\ $w=6$),
Fig. \ref{spreading2D} compares our spreading methods \textbf{GM-sort} and \textbf{SM} against
the baseline method \textbf{GM}, in 2D (top), 3D (bottom), and for ``rand'' 
(left) and ``cluster'' (right) distributions.
Solid lines show total times (in nanoseconds per nonuniform point)
including preprocessing (sorting) of new nonuniform points.
Dotted lines show execution excluding this, so are relevant for {\em repeated transforms} with same set of nonuniform points.

We can see from the ``rand'' results that for large grids
($n_1=n_2\geq 2^{10}$ in 2D, or $n_1=n_2=n_3 \geq 2^7$ in 3D) bin-sorting
brings a large gain.
In the extreme case (the largest $n_i$ tested), \textbf{GM-sort} is 3.9$\times$ 
faster than \textbf{GM} in 2D, and 7.6$\times$ faster in 3D.
%The greater speedup in 3D is because the data is more spread out. % speculative
%We expect to
%see greater speedup in 2D as well if tested with larger $n_i$.
On the other hand, for small grids, because the memory accesses
are already localized, we do not see any benefit of bin-sorting.

From the ``cluster'' results, since nonuniform points all reside
in a small region, bin-sorting brings no benefit.
However, we see the
clear advantage of doing local spreading on shared memory, in that \textbf{SM} is
up to 12.8$\times$ faster than \textbf{GM} in 2D, and up to 3.2$\times$ faster
in 3D.  
The speedup in 3D is limited because the padding of the bins,
especially in the $z$ direction,
grows the volume of global atomic adds needed in {\em Step 3}.

Comparing the dark blue curves in the left and right plots, we see a
distribution-robust performance for \textbf{SM}, in that
similar throughput is achieved for ``rand'' and ``cluster'' distributions.  
In comparison, \textbf{GM-sort} is on average 3.9$\times$ slower in 2D 
comparing ``cluster'' and ``rand'' distributions.
The 2D execution throughput (excluding precomputation) exceeds
$10^9$ points/sec for large grids, and is close even when including precomputation.

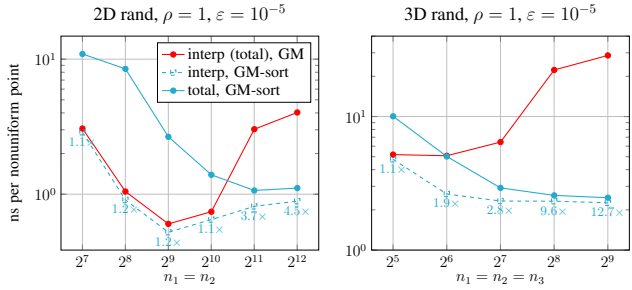
\begin{figure}[t]  % fffffffffffffffffffffffffffffffffffffffffffffffffffffff
  \centering
  % 4 plots here

\begin{tikzpicture}[scale=0.5]
	\begin{groupplot}[group style={{group size=2 by 3},
		vertical sep=6em, 
		horizontal sep=4em},
		log basis x={2},
		grid,
		legend pos=north east,
		legend cell align=left,
		ytick={1,10,100},
		ymode=log,
		xmode=log,
		every node near coord/.append style={
			/pgf/number format/fixed,
			/pgf/number format/precision=1,}
		]
		%% rand, rho=1
		\nextgroupplot[title={2D rand, $\rho=1$, $\varepsilon=10^{-5}$},
		ylabel=ns per nonuniform point,
		xlabel={$n_1=n_2$}]
		\addplot +[discard if not={method}{0},
		discard if not={density}{1},
		discard if ={nf1}{32},
		discard if ={nf1}{64},
		col-nu, solid, thick,
		mark=*, mark options={fill=col-nu}]
		table [x=nf1,y expr={1e9*1e-3*(\thisrow{spread})/\thisrow{M}}]{data/interp2d_density_uniform_withspeedup.dat};
		\addlegendentry{interp (total), {GM}}
		\addplot +[discard if not={method}{1},
		discard if not={density}{1},
		discard if ={nf1}{32},
		discard if ={nf1}{64},
		col-sb-2, dashed, thick,
		mark=square,mark options={fill=col-sb-2},
		point meta=explicit,
		nodes near coords=\pgfmathprintnumber{\pgfplotspointmeta}$\times$,
		nodes near coords align={below},]
		table [meta=speedup,x=nf1,y expr={1e9*1e-3*(\thisrow{spread})/\thisrow{M})}]{data/interp2d_density_uniform_withspeedup.dat};
		\addlegendentry{interp, {GM-sort}}
		\addplot +[discard if not={method}{1},
		discard if not={density}{1},
		discard if ={nf1}{32},
		discard if ={nf1}{64},
		col-sb-2, thick,
		mark=*,mark options={fill=col-sb-2}]
		table [x=nf1,y expr={1e9*1e-3*(\thisrow{spreadprop}+\thisrow{spread})/\thisrow{M}}]{data/interp2d_density_uniform_withspeedup.dat};
		\addlegendentry{total, {GM-sort}}

		%% cluster, rho=1
		%\nextgroupplot[title={2D cluster, $\rho=1$, $\varepsilon=10^{-5}$},
		%xlabel={$n_1=n_2$}]
		%\addplot +[discard if not={method}{0},
		%discard if not={density}{1},
		%discard if ={nf1}{32},
		%discard if ={nf1}{64},
		%col-nu, solid, thick,
		%mark=*, mark options={fill=col-nu}]
		%table [x=nf1,y expr={1e9*1e-3*(\thisrow{spread})/\thisrow{M}}]{data/interp2d_density_cluster_withspeedup.dat};
		%\addplot +[discard if not={method}{1},
		%discard if not={density}{1},
		%discard if ={nf1}{32},
		%discard if ={nf1}{64},
		%col-sb-2, dashed, thick,
		%mark=square,mark options={fill=col-sb-2},
		%point meta=explicit,
		%nodes near coords=\pgfmathprintnumber{\pgfplotspointmeta}$\times$,
		%nodes near coords align={below},]
		%table [meta=speedup,x=nf1,y expr={1e9*1e-3*(\thisrow{spread})/\thisrow{M})}]{data/interp2d_density_cluster_withspeedup.dat};
		%\addplot +[discard if not={method}{1},
		%discard if not={density}{1},
		%discard if ={nf1}{32},
		%discard if ={nf1}{64},
		%col-sb-2, thick,
		%mark=*,mark options={fill=col-sb-2}]
		%table [x=nf1,y expr={1e9*1e-3*(\thisrow{spreadprop}+\thisrow{spread})/\thisrow{M}}]{data/interp2d_density_cluster_withspeedup.dat};

		%% 3D rand, rho=1
		\nextgroupplot[title={3D rand, $\rho=1$, $\varepsilon=10^{-5}$},
		xlabel={$n_1=n_2=n_3$},
		ymin=1e9*1e-9,
		ymax=1e9*4e-8,
		]
		\addplot +[discard if not={method}{0},
		discard if not={density}{1.0},
		col-nu, solid, thick,
		mark=*, mark options={fill=col-nu}]
		table [x=nf1,y expr={1e9*1e-3*(\thisrow{spread})/\thisrow{M}}]{data/interp3d_density_uniform_withspeedup.dat};
		\addplot +[discard if not={method}{1},
		discard if not={density}{1.0},
		col-sb-2, dashed, thick,
		mark=square,mark options={fill=col-sb-2},
		point meta=explicit,
		nodes near coords=\pgfmathprintnumber{\pgfplotspointmeta}$\times$,
		nodes near coords align={below},]
		table [meta=speedup,x=nf1,y expr={1e9*1e-3*(\thisrow{spread})/\thisrow{M})}]{data/interp3d_density_uniform_withspeedup.dat};
		\addplot +[discard if not={method}{1},
		discard if not={density}{1.0},
		col-sb-2, thick,
		mark=*,mark options={fill=col-sb-2}]
		table [x=nf1,y expr={1e9*1e-3*(\thisrow{spreadprop}+\thisrow{spread})/\thisrow{M}}]{data/interp3d_density_uniform_withspeedup.dat};

		%% 3D cluster, rho=1
		%\nextgroupplot[title={3D cluster, $\rho=1$, $\varepsilon=10^{-5}$},
		%xlabel={$n_1=n_2=n_3$},
		%ymin=1e9*1e-9,
		%ymax=1e9*2e-8,]
		%\addplot +[discard if not={method}{0},
		%discard if not={density}{1.0},
		%col-nu, solid, thick,
		%mark=*, mark options={fill=col-nu}]
		%table [x=nf1,y expr={1e9*1e-3*(\thisrow{spread})/\thisrow{M}}]{data/interp3d_density_cluster_withspeedup.dat};
		%\addplot +[discard if not={method}{1},
		%discard if not={density}{1.0},
		%col-sb-2, dashed, thick,
		%mark=square,mark options={fill=col-sb-2},
		%point meta=explicit,
		%nodes near coords=\pgfmathprintnumber{\pgfplotspointmeta}$\times$,
		%nodes near coords align={below},]
		%table [meta=speedup,x=nf1,y expr={1e9*1e-3*(\thisrow{spread})/\thisrow{M})}]{data/interp3d_density_cluster_withspeedup.dat};
		%\addplot +[discard if not={method}{1},
		%discard if not={density}{1.0},
		%col-sb-2, thick,
		%mark=*,mark options={fill=col-sb-2}]
		%table [x=nf1,y expr={1e9*1e-3*(\thisrow{spreadprop}+\thisrow{spread})/\thisrow{M}}]{data/interp3d_density_cluster_withspeedup.dat};
	\end{groupplot}
\end{tikzpicture}
  \caption{Interpolation comparisons. Execution time per
    nonuniform point is shown, for various fine grid sizes, and distribution ``rand'', 
	in 2D and 3D.
    The ``total'' time (solid lines) includes the bin-sorting precomputation,
    whereas the ``interp'' time (dotted lines)
    excludes this precomputation.}
  \label{interpolation2D}
\end{figure}

% IIIIIIIIIIIIIIIIIIIIIIIIIIIIIIIIIIIIIIIIIIIIIIIIIIIIIIIIIIIIIIIIIIIIIIIIIII
\subsection{Interpolation performance}
%\msnote{
%Similar as \textbf{GM-sort} vs \textbf{GM} in spreading, bin-sorting helps for 
%large find grid. The difference is that for interpolation, since there
%is no write confliction, the ``interp'' time of \textbf{GM-sort} doesn't become slower 
%than GM.
%}
For the same accuracy,
Fig. \ref{interpolation2D} compares interpolation method \textbf{GM-sort} against \textbf{GM} in 
2D and 3D, for the ``rand'' distribution.
(We exclude the ``cluster'' results, since, as with spreading, bin-sorting nonuniform points has no effect.)
We see that again \textbf{GM-sort}
improves the performance for large grids ($n_1=n_2\geq 2^{11}$ in 2D,
or $n_1=n_2=n_3\geq 2^6$ in 3D).
It is 4.5$\times$ faster than \textbf{GM} in 2D, and 12.7$\times$ faster in 3D, for the largest $n_i$ tested. 
A difference with spreading is that, because there are no
global write conflicts,
the execution time of \textbf{GM-sort} (excluding precomputation)
never becomes slower than \textbf{GM}.

% Do we want this?
%\begin{figure}[ht]   % fffffffffffffffffffffffffffffffffffffffffffffffffffffff
%  \centering
%  \input{fig/compacc}
%  \caption{Comparison of empirical relative error versus kernel width
%    $w=3,5,7$ from different libraries. 
%	%The libraries are set up as
%    %described in Sec. \ref{sec:benchmarkwithothercode}. 
%	In each case, $N_1=N_2(=N_3)$ and $M=\prod_{i=1}^d N_i$. Note that in
%	general, smaller relative errors are obtained with larger kernel width. For
%	gpuNUFFT, the errors stay approximately the same for
%    $w\geq5$.}
%  \label{fig:acc}
%\end{figure}

% CCCCCCCCCCCCCCCCCCCCCCCCCCCCCCCCCCCCCCCCCCCCCCCCCCCCCCCCCCCCCCCCCCCCCCCCCCCC
\subsection{Benchmark comparisons against existing libraries}
\label{sec:bc}

We now compare {cuFINUFFT} against the CPU library {FINUFFT}
(which has already benchmarked favorably against other CPU libraries \cite{Barnett_2019}),
and GPU libraries {CUNFFT} \cite{cunfft} and {gpuNUFFT} \cite{Strelak2019}.
For {FINUFFT}, we used a high-end compute node equipped with 512 GB RAM and
two Intel Xeon E5-2680 v4 processors (released in 2016).  Each processor has 14 physical cores at 2.40 GHz. We ran multithreaded {FINUFFT} with 28 threads (1 thread per physical core).

\begin{itemize}
\item {cuFINUFFT} version 1.0. We used host compiler flags \texttt{-fPIC -O3 
	  -funroll-loops -march=native}.   % -g  !
\item {FINUFFT} version 2.0.2. Compiler flags were \texttt{-O3 -funroll-loops -march=native 
	  -fcx-limited-range}. We fixed upsampling factor $\sigma=2$
	  to match that used in cuFINUFFT.
\item {CUNFFT} version 2.0. We compiled with
	  fast Gaussian gridding (\texttt{-DCOM\_FG\_PSI=ON}).
	  %To use kernel width $w$, we compile the code 
	  %with cut-off parameter $m$ (\texttt{-DCUT\_OFF=m}) such that $w=2m+1$. 
	  Default dimensions of thread blocks (\texttt{THREAD\_DIM\_X=16}, 
	  \texttt{THREAD\_DIM\_Y=16}) are used. %\url{https://github.com/sukunis/CUNFFT} 
        \item {gpuNUFFT}
          version 2.1.0. We use its MATLAB interface. 
	  We used default host compiler flags
	  (\texttt{-std=c++11 -fPIC}). We set \texttt{MAXIMUM\_ALIASING\_ERROR} to 
	  $10^{-6}$ to get more accurate results.
	  %We run the code with kernel width $w$.
	  We use the same sector width $8$
          and \texttt{THREAD\_BLOCK\_SIZE=256} as the demo codes.
	  %\url{https://github.com/andyschwarzl/gpuNUFFT}
\end{itemize}

We present three different timings for NUFFT executions:
\begin{itemize}
\item ``total'': Execution time (per nonuniform point) %from fresh inputs,
  for inputs and output on the GPU.
  %GPU arrays $x, y, z, c~(\tilde{c}),
  %\tilde{f}~(f)$ are given as both inputs and outputs.  
\item ``total+mem'': Execution time (per nonuniform point), including the time for GPU memory allocation plus transferring data from host to GPU and back.
\item ``exec'': Execution time (per nonuniform point) for a transform, after its nonuniform points have already been preprocessed. This is a subset of
  the ``total'' time. It is the relevant time for the case of multiple fixed-size transforms with a fixed set of nonuniform points, but new strength or coefficient vectors.
\end{itemize}

There is a constant start-up cost (about 0.1--0.2 second) for calling 
the {cuFFT} library, so to exclude it we add a dummy call of
\texttt{cuFFTPlan1d} before calling {cuFINUFFT} or {CUNFFT}. For 
gpuNUFFT, in ``total+mem'', we exclude the time for building the
nonuniform FFT operator and creating the cuFFT plan.
Note also that gpuNUFFT sorts the nonuniform points into
sectors on the CPU and copies the arrays to the GPU when it builds the
operator, so, to be generous, we do not include this in ``total+mem''
either. Finally, ``total'' is not shown for {gpuNUFFT} and {CUNFFT}, because
{gpuNUFFT} takes CPU arrays as inputs and outputs, and
{CUNFFT} allocates GPU memory in the initialization stage (\texttt{cunfft\_init})
in a way that did not allow us to separate its timing.

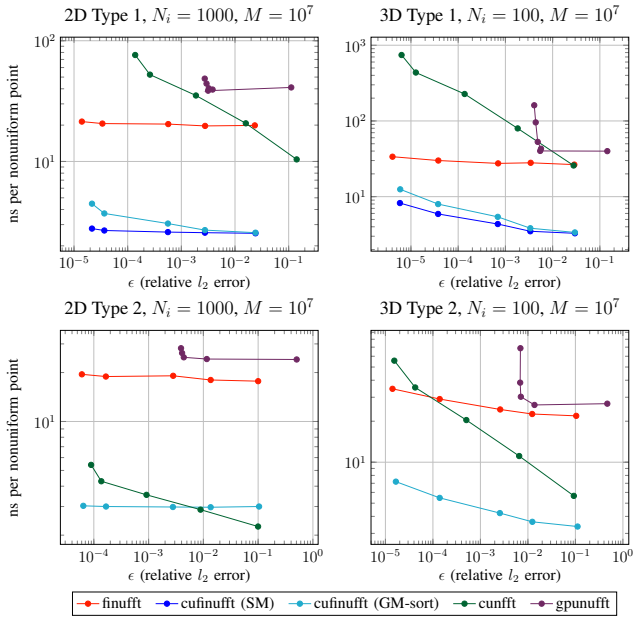
\begin{figure}[ht]  % fffffffffffffffffffffffffffffffffffffffffffffffffffff
  \centering
  % 4 plots
\begin{tikzpicture}[scale=0.5,transform shape]
\begin{groupplot}[group style={{group size=2 by 2},
	                            vertical sep=6em,
								horizontal sep=4em},
				  log basis x={10},
				  grid,
  	              legend cell align=left,
				  legend style={/tikz/every even column/.append style={column sep=0.3cm}},
				  xlabel=$\epsilon$ (relative $l_2$ error),
				  ytick={1,10,100,1000},
				  ymode=log,
				  xmode=log]
	\nextgroupplot[title={2D Type 1, $N_i=1000$, $M=10^7$},
	               ylabel=ns per nonuniform point,
				   legend to name={commonlegend-t1t},
				   legend style={legend columns=5},
				   ymode=log,
				   xmode=log]
	%\draw[{<[scale=2.5,length=2,width=3]}-{>[scale=2.5,length=2,width=3]}]({axis cs:1e-2,0}|-{axis cs:0,3})--({axis cs:1e-2,0}|-{axis cs:0,15})node[midway,xshift=10pt]{$5\times$};
	%\draw[{<[scale=2.5,length=2,width=3]}-{>[scale=2.5,length=2,width=3]}]({axis cs:1e-4,0}|-{axis cs:0,4})--({axis cs:1e-4,0}|-{axis cs:0,60})node[midway,xshift=10pt]{$15\times$};
    \addplot +[discard if not={nupts}{1},
	           discard if not={code}{0},
               col-finufft,
               mark=*,mark options={fill=col-finufft}]
               table [x=acc,y expr={1e9*(\thisrow{total})/\thisrow{M}}]{\compdata/2d1_acc_cpu.dat};
	\addlegendentry{finufft}
    \addplot +[discard if not={nupts}{1},
    		   discard if not={code}{2},
    		   col-cufinufft,
    		   mark=*,mark options={fill=col-cufinufft}]
    		  table [x=acc,y expr={1e9*\thisrow{totalgpumem}/\thisrow{M}}]{\compdata/2d1_acc.dat};
	\addlegendentry{cufinufft (SM)}
    \addplot +[discard if not={nupts}{1},
    		   discard if not={code}{1},
    		   col-cufinufft-2, solid, 
    		   mark=*,mark options={fill=col-cufinufft-2}]
    		  table [x=acc,y expr={1e9*\thisrow{totalgpumem}/\thisrow{M}}]{\compdata/2d1_acc.dat};
	\addlegendentry{cufinufft (GM-sort)}
    \addplot +[discard if not={nupts}{1},
			   discard if not={code}{3},
			   col-cunfft, solid,
			   mark=*,mark options={fill=col-cunfft}]
			   table [x=acc,y expr={1e9*\thisrow{totalgpumem}/\thisrow{M}}]{\compdata/2d1_acc.dat};
	\addlegendentry{cunfft}
    \addplot +[discard if not={nupts}{1},
			  discard if not={code}{4},
			  col-gpunufft, solid,
			  mark=*,mark options={fill=col-gpunufft}]
    		  table [x=acc,y expr={1e9*\thisrow{totalgpumem}/\thisrow{M}}]{\compdata/2d1_gpunufft_acc.dat};
	\addlegendentry{gpunufft}

	\nextgroupplot[title={3D Type 1, $N_i=100$, $M=10^7$},]
	%\draw[{<[scale=2.5,length=2,width=3]}-{>[scale=2.5,length=2,width=3]}]({axis cs:1e-2,0}|-{axis cs:0,5})--({axis cs:1e-2,0}|-{axis cs:0,25})node[midway,xshift=10pt]{$5\times$};
    \addplot +[discard if not={nupts}{1},
	           discard if not={code}{0},
               col-finufft,
               mark=*,mark options={fill=col-finufft}]
               table [x=acc,y expr={1e9*(\thisrow{total})/\thisrow{M}}]{\compdata/3d1_acc_cpu.dat};
    \addplot +[discard if not={nupts}{1},
			   discard if not={code}{2},
    		   col-cufinufft,
    		   mark=*,mark options={fill=col-cufinufft}]
    		   table [x=acc,y expr={1e9*\thisrow{totalgpumem}/\thisrow{M}}]{\compdata/3d1_acc.dat};
    \addplot +[discard if not={nupts}{1},
    		   discard if not={code}{1},
    		   col-cufinufft-2, solid,
    		   mark=*,mark options={fill=col-cufinufft-2}]
    		  table [x=acc,y expr={1e9*\thisrow{totalgpumem}/\thisrow{M}}]{\compdata/3d1_acc.dat};
    \addplot +[discard if not={nupts}{1},
			   discard if not={code}{3},
			   col-cunfft, solid,
			   mark=*,mark options={fill=col-cunfft}]
			   table [x=acc,y expr={1e9*\thisrow{totalgpumem}/\thisrow{M}}]{\compdata/3d1_acc.dat};
   \addplot +[discard if not={nupts}{1},
			  discard if not={code}{4},
			  col-gpunufft, solid,
			  mark=*,mark options={fill=col-gpunufft}]
    		  table [x=acc,y expr={1e9*\thisrow{totalgpumem}/\thisrow{M}}]{\compdata/3d1_gpunufft_acc.dat};

	\nextgroupplot[title={2D Type 2, $N_i=1000$, $M=10^7$},
	               ylabel=ns per nonuniform point,]
	%\draw[{<[scale=2.5,length=2,width=3]}-{>[scale=2.5,length=2,width=3]}]({axis cs:5e-2,0}|-{axis cs:0,3.1})--({axis cs:5e-2,0}|-{axis cs:0,15.5})node[midway,xshift=10pt]{$5\times$};
    \addplot +[discard if not={nupts}{1},
	           discard if not={code}{0},
               col-finufft,
               mark=*,mark options={fill=col-finufft}]
               table [x=acc,y expr={1e9*(\thisrow{total})/\thisrow{M}}]{\compdata/2d2_acc_cpu.dat};
    \addplot +[discard if not={nupts}{1},
    		   discard if not={code}{1},
    		   col-cufinufft-2,
    		   mark=*,mark options={fill=col-cufinufft-2}]
    		  table [x=acc,y expr={1e9*\thisrow{totalgpumem}/\thisrow{M}}]{\compdata/2d2_acc.dat};
    \addplot +[discard if not={nupts}{1},
			   discard if not={code}{3},
			   col-cunfft, solid,
			   mark=*,mark options={fill=col-cunfft}]
			   table [x=acc,y expr={1e9*\thisrow{totalgpumem}/\thisrow{M}}]{\compdata/2d2_acc.dat};
    \addplot +[discard if not={nupts}{1},
			  discard if not={code}{4},
			  col-gpunufft, solid,
			  mark=*,mark options={fill=col-gpunufft}]
    		  table [x=acc,y expr={1e9*\thisrow{totalgpumem}/\thisrow{M}}]{\compdata/2d2_gpunufft_acc.dat};

	\nextgroupplot[title={3D Type 2, $N_i=100$, $M=10^7$},]
	%\draw[{<[scale=2.5,length=2,width=3]}-{>[scale=2.5,length=2,width=3]}]({axis cs:0.2,0}|-{axis cs:0,4})--({axis cs:0.2,0}|-{axis cs:0,20})node[midway,xshift=10pt]{$5\times$};
    \addplot +[discard if not={nupts}{1},
	           discard if not={code}{0},
               col-finufft,
               mark=*,mark options={fill=col-finufft}]
               table [x=acc,y expr={1e9*(\thisrow{total})/\thisrow{M}}]{\compdata/3d2_acc_cpu.dat};
    \addplot +[discard if not={nupts}{1},
    		   discard if not={code}{1},
    		   col-cufinufft-2,
    		   mark=*,mark options={fill=col-cufinufft-2}]
    		  table [x=acc,y expr={1e9*\thisrow{totalgpumem}/\thisrow{M}}]{\compdata/3d2_acc.dat};
    \addplot +[discard if not={nupts}{1},
			   discard if not={code}{3},
			   col-cunfft, solid,
			   mark=*,mark options={fill=col-cunfft}]
			   table [x=acc,y expr={1e9*\thisrow{totalgpumem}/\thisrow{M}}]{\compdata/3d2_acc.dat};
    \addplot +[discard if not={nupts}{1},
			  discard if not={code}{4},
			  col-gpunufft, solid,
			  mark=*,mark options={fill=col-gpunufft}]
    		  table [x=acc,y expr={1e9*\thisrow{totalgpumem}/\thisrow{M}}]{\compdata/3d2_gpunufft_acc.dat};
\end{groupplot}
\path (group c1r2.west|-current bounding box.south) -- node[below]{\ref{commonlegend-t1t}} (group c2r2.east|-current bounding box.south);
\end{tikzpicture}
	\caption{Single precision NUFFT comparisons in 2D (left) and 3D 
	(right), for type 1 (upper) and 2 (lower).  ``total+mem'' (``total'' for FINUFFT) time per nonuniform 
	point vs accuracy is shown, for the named libraries, for the distribution ``rand''.
	}
  \label{fig:single}
\end{figure}

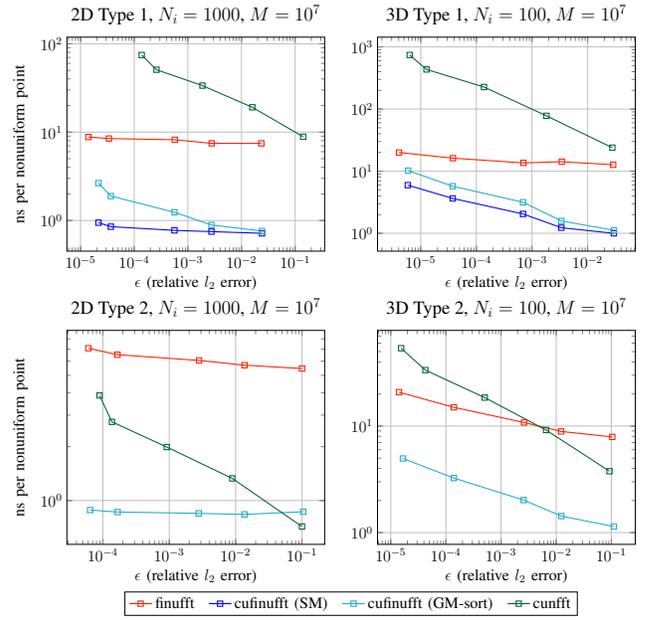
\begin{figure}[ht]  % fffffffffffffffffffffffffffffffffffffffffffffffffffff
  \centering
  % 4 plots
\begin{tikzpicture}[scale=0.5,transform shape]
\begin{groupplot}[group style={{group size=2 by 2},
	                            vertical sep=6em,
								horizontal sep=4em},
				  log basis x={10},
				  grid,
  	              legend cell align=left,
				  legend style={/tikz/every even column/.append style={column sep=0.3cm}},
				  xlabel=$\epsilon$ (relative $l_2$ error),
				  ytick={1,10,100,1000},
				  ymode=log,
				  xmode=log]
	\nextgroupplot[title={2D Type 1, $N_i=1000$, $M=10^7$},
	               ylabel=ns per nonuniform point,
				   legend to name={commonlegend-t1e},
				   legend style={legend columns=4},
				   ymode=log,
				   xmode=log]
	%\draw[{<[scale=2.5,length=2,width=3]}-{>[scale=2.5,length=2,width=3]}]({axis cs:1e-2,0}|-{axis cs:0,1})--({axis cs:1e-2,0}|-{axis cs:0,15})node[midway,xshift=10pt]{$15\times$};
	%\draw[{<[scale=2.5,length=2,width=3]}-{>[scale=2.5,length=2,width=3]}]({axis cs:5e-5,0}|-{axis cs:0,1})--({axis cs:5e-5,0}|-{axis cs:0,80})node[pos=0.6,xshift=12pt]{$80\times$};
    \addplot +[discard if not={nupts}{1},
	           discard if not={code}{0},
               col-finufft, solid,
               mark=square,mark options={fill=col-finufft},thick]
               table [x=acc,y expr={1e9*(\thisrow{exec})/\thisrow{M}}]{\compdata/2d1_acc_cpu.dat};
	\addlegendentry{finufft}
    \addplot +[discard if not={nupts}{1},
              discard if not={code}{2},
              col-cufinufft, solid,
              mark=square, mark options={solid, fill=col-cufinufft},
              thick]
              table [x=acc,y expr={1e9*(\thisrow{exec})/\thisrow{M}}]{\compdata/2d1_acc.dat};
	\addlegendentry{cufinufft (SM)}
    \addplot +[discard if not={nupts}{1},
              discard if not={code}{1},
              col-cufinufft-2, solid,
              mark=square, mark options={solid, fill=col-cufinufft!20!black},
              thick]
              table [x=acc,y expr={1e9*(\thisrow{exec})/\thisrow{M}}]{\compdata/2d1_acc.dat};
	\addlegendentry{cufinufft (GM-sort)}
    \addplot +[discard if not={nupts}{1},
              discard if not={code}{3},
              col-cunfft, solid,
              mark=square,mark options={solid, fill=col-cunfft},
              thick]
              table [x=acc,y expr={1e9*(\thisrow{exec})/\thisrow{M}}]{\compdata/2d1_acc.dat};
	\addlegendentry{cunfft}
	\nextgroupplot[title={3D Type 1, $N_i=100$, $M=10^7$},]
	%\draw[{<[scale=2.5,length=2,width=3]}-{>[scale=2.5,length=2,width=3]}]({axis cs:1e-2,0}|-{axis cs:0,1.5})--({axis cs:1e-2,0}|-{axis cs:0,22.5})node[midway,xshift=10pt]{$15\times$};
    \addplot +[discard if not={nupts}{1},
	           discard if not={code}{0},
               col-finufft, solid,
               mark=square,mark options={fill=col-finufft},thick]
               table [x=acc,y expr={1e9*(\thisrow{exec})/\thisrow{M}}]{\compdata/3d1_acc_cpu.dat};
    \addplot +[discard if not={nupts}{1},
              discard if not={code}{2},
              col-cufinufft, solid,
              mark=square, mark options={solid, fill=col-cufinufft},
              thick]
              table [x=acc,y expr={1e9*(\thisrow{exec})/\thisrow{M}}]{\compdata/3d1_acc.dat};
    \addplot +[discard if not={nupts}{1},
              discard if not={code}{1},
              col-cufinufft-2, solid,
              mark=square, mark options={solid, fill=col-cufinufft-2},
              thick]
              table [x=acc,y expr={1e9*(\thisrow{exec})/\thisrow{M}}]{\compdata/3d1_acc.dat};
    \addplot +[discard if not={nupts}{1},
              discard if not={code}{3},
              col-cunfft, solid,
              mark=square,mark options={solid, fill=col-cunfft},
              thick]
              table [x=acc,y expr={1e9*(\thisrow{exec})/\thisrow{M}}]{\compdata/3d1_acc.dat};
	\nextgroupplot[title={2D Type 2, $N_i=1000$, $M=10^7$},
	               ylabel=ns per nonuniform point,]
	%\draw[{<[scale=2.5,length=2,width=3]}-{>[scale=2.5,length=2,width=3]}]({axis cs:5e-2,0}|-{axis cs:0,1})--({axis cs:5e-2,0}|-{axis cs:0,5})node[midway,xshift=10pt]{$5\times$};
    \addplot +[discard if not={nupts}{1},
	           discard if not={code}{0},
               col-finufft, solid,
               mark=square,mark options={fill=col-finufft},thick]
               table [x=acc,y expr={1e9*(\thisrow{exec})/\thisrow{M}}]{\compdata/2d2_acc_cpu.dat};
    \addplot +[discard if not={nupts}{1},
              discard if not={code}{1},
              col-cufinufft-2, solid,
              mark=square, mark options={solid, fill=col-cufinufft-2},
              thick]
              table [x=acc,y expr={1e9*(\thisrow{exec})/\thisrow{M}}]{\compdata/2d2_acc.dat};
    \addplot +[discard if not={nupts}{1},
              discard if not={code}{3},
              col-cunfft, solid,
              mark=square,mark options={solid, fill=col-cunfft},
              thick]
              table [x=acc,y expr={1e9*(\thisrow{exec})/\thisrow{M}}]{\compdata/2d2_acc.dat};
	\nextgroupplot[title={3D Type 2, $N_i=100$, $M=10^7$},]
	%\draw[{<[scale=2.5,length=2,width=3]}-{>[scale=2.5,length=2,width=3]}]({axis cs:1e-3,0}|-{axis cs:0,2.3})--({axis cs:1e-3,0}|-{axis cs:0,11.5})node[midway,xshift=10pt]{$5\times$};
    \addplot +[discard if not={nupts}{1},
	           discard if not={code}{0},
               col-finufft, solid,
               mark=square,mark options={fill=col-finufft},thick]
               table [x=acc,y expr={1e9*(\thisrow{exec})/\thisrow{M}}]{\compdata/3d2_acc_cpu.dat};
    \addplot +[discard if not={nupts}{1},
              discard if not={code}{1},
              col-cufinufft-2, solid,
              mark=square, mark options={solid, fill=col-cufinufft-2},
              %point meta=explicit,
              %nodes near coords=\pgfmathprintnumber{\pgfplotspointmeta}x,
              %nodes near coords align={below},
              thick]
              %table [meta=speedupexec,x=acc,y expr={1e9*(\thisrow{exec})/\thisrow{M}}]{\compdata/3d1_double.dat};
              table [x=acc,y expr={1e9*(\thisrow{exec})/\thisrow{M}}]{\compdata/3d2_acc.dat};
    \addplot +[discard if not={nupts}{1},
              discard if not={code}{3},
              col-cunfft, solid,
              mark=square,mark options={solid, fill=col-cunfft},
              thick]
              table [x=acc,y expr={1e9*(\thisrow{exec})/\thisrow{M}}]{\compdata/3d2_acc.dat};
\end{groupplot}
\path (group c1r2.west|-current bounding box.south) -- node[below]{\ref{commonlegend-t1e}} (group c2r2.east|-current bounding box.south);
\end{tikzpicture}
	\caption{Single precision comparisons in 2D and 3D. ``exec''
	time per nonuniform point vs accuracy is shown for the tested libraries,
	except for gpuNUFFT. 
	For explanation see caption of Fig. \ref{fig:single}.
	}
  \label{fig:singleexec}
\end{figure}

\begin{figure}[ht]   % ffffffffffffffffffffffffffffffffffffffffffffffffffffffff
  \centering
  \input{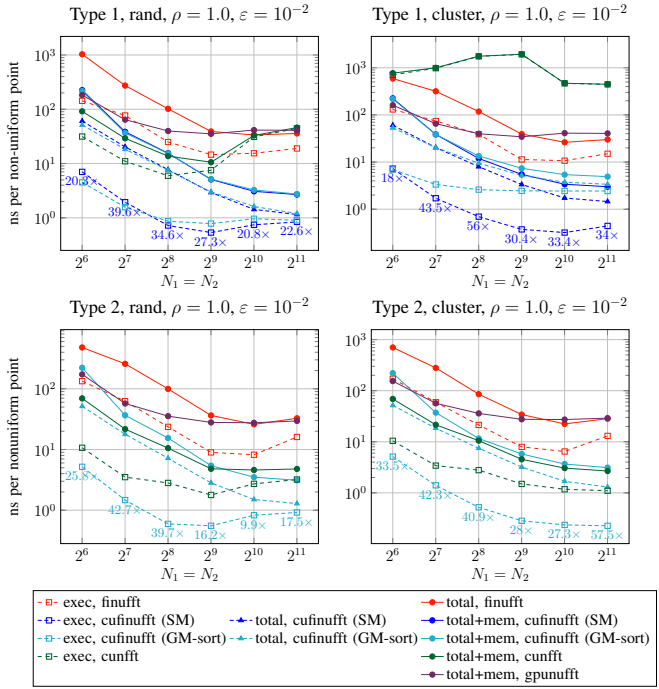}
	\caption{Detailed 2D Type 1 (top) and 2 (bottom) NUFFT comparisons (single precision).
	Execution time per nonuniform point vs number of Fourier modes are shown
	for the named libraries, comparing ``rand'' (left) and ``cluster'' (right).
	Annotations give the speedup of ``exec'' of {cuFINUFFT} (SM) for type 1, 
	{cuFINUFFT} (GM-sort) for type 2, vs ``exec'' of {FINUFFT}.
	}
  \label{fig:type12}
\end{figure}

\begin{figure}[ht]  % fffffffffffffffffffffffffffffffffffffffffffffffffffff
  \centering
  \input{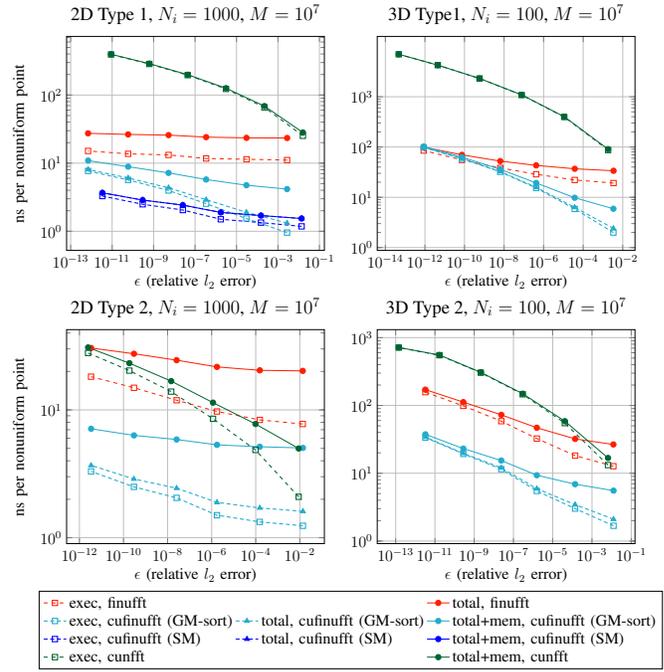}
  \caption{Double precision comparisons. All three timings ``exec'',
	``total'', and ``total+mem'' are shown. For more 
	explanation see caption of Fig.~\ref{fig:single}.}
  \label{fig:double}
\end{figure}

We now discuss the results (\Cref{fig:single,fig:singleexec,fig:double,fig:type12} and
Table \ref{table:3dtype1}).
A wide range of relative $\ell_2$ errors, $\epsilon$, are explored by varying
the requested tolerance, or kernel parameters (usually the width $w$), for each library.
Error is measured against a ground truth of FINUFFT with tolerance
$\varepsilon=10^{-14}$ for double precision runs, and $\varepsilon=6\times
10^{-8}$ for single precision runs.

\begin{table*}[ht]  % ttttttttttttttttttttttttttttttttttttttttttttttttttttttttt
  \centering
  \pgfplotstableread[col sep=comma]{\compdata/3d1_sepcufinufft.dat}{\table}
  \pgfkeys{
	/pgf/number format/.cd,
	sci,
	sci generic={mantissa sep=\times,exponent={10^{#1}}}}
  \pgfplotstabletypeset[
	every head row/.style={
		before row={
			\toprule
		},
		after row=\midrule},
	every last row/.style={after row=\bottomrule},
    %fixed zerofill,     % Fill numbers with zeros
	every nth row={4}{before row=\midrule},
	columns={tol,N1,M,code,exec,ram,
	         speedupoverall,spreadpercent},
	columns/code/.style={column type=l,string type,column name=Method},
	columns/tol/.style={string type,column name=$\varepsilon$},
	columns/N1/.style={int detect, column name={$N_1=N_2=N_3$}},
	columns/spreadpercent/.style={fixed,column name=Spread fraction (\%), precision=1},
	columns/exec/.style={column type=c,column name=``Exec'' time (sec),
				%std=-3:3,
				fixed,
				%fixed zerofill=false,
                %sci,sci zerofill,   % fixed can't fix 2 sig figs, matlab %.2g !
                precision=4},
	columns/ram/.style={set thousands separator={},int detect, column type=c,column name=RAM (MB)},
	columns/speedupoverall/.style={fixed,column name=Speedup vs FINUFFT,
		postproc cell content/.append style={
		/pgfplots/table/@cell content/.add={$\bf}{$},
		},postproc cell content/.append style={
		/pgfplots/table/@cell content/.add={}{$\times$},
		},precision=1},
	multicolumn names=c,
    ] {\table}
\caption{{cuFINUFFT} 3D type 1 NUFFT GPU memory usage, and ``exec'' time, for 
	distribution ``rand'' and for two relative tolerances
	$\varepsilon=10^{-2}, 10^{-5}$. 
	Speedup is computed relative to the ``exec'' time from {FINUFFT}. 
	Spread fraction is the percentage of ``exec'' time spent on spreading.
	RAM is measured using \texttt{nvidia-smi}. For the baseline spreading
	method \textbf{GM}, RAM use is 381 MB for
	$N_i=32$ and 5113 MB for $N_i=256$).}
  \label{table:3dtype1}
\end{table*}

% SSSSSSSSSSSSSSSSSSSSSSSSSSSSSSSSSSSSSSSSSSSSSSSSSSSSSSSSSSSSSSSSSSSSSSSSSSSSS
\paragraph{Single precision comparisons}
%fig:single
Figs.~\ref{fig:single} and \ref{fig:singleexec} compare performance of both type 1
(top) and type 2 (bottom) in 2D (left), 3D
(right), for all libraries in single precision. 
We can see from the top plots that for type 1, {cuFINUFFT} outperforms
all other libraries. For type 1, the best performance is achieved using the \textbf{SM} method (dark blue).
The ``exec'' time of {cuFINUFFT ({SM})} in 2D is
around 10$\times$ faster than ``exec'' time of {FINUFFT}, independent of the
accuracy; and in 3D, it is 3--12$\times$ faster from high to low accuracy. 

For type 2 (bottom plots),
except for 2D low accuracy ($\epsilon>10^{-2}$)
where CUNFFT is comparable to {cuFINUFFT}, 
{cuFINUFFT} is again the fastest.
Its ``exec'' time is 4--7$\times$ and 6--8$\times$ faster than the ``exec''
time of {FINUFFT} in 2D and 3D respectively.

In Fig. \ref{fig:type12}, we fix a tolerance
$\varepsilon=10^{-2}$ (achievable by all libraries),
and examine the effect of 
nonuniform point distribution and
number of Fourier modes (fixing density $\rho=1$) on library performance.
%fig:type1top
From the top plots, for type 1, we observe distribution-robust
performance in {cuFINUFFT} ({SM}), {FINUFFT} and {gpuNUFFT}. 
The others, {cuFINUFFT} ({GM-sort}) 
and {CUNFFT}, slow down when the points are clustered: for an intermediate
problem
size ($N_i=2^9$), {cuFINUFFT} ({GM-sort}) ``exec''
is slowed by a factor of 3 when switching from
``rand'' to ``cluster'' to ``rand''. Dramatically,
{CUNFFT} is slowed by a factor of 200: it is very slow for clustered
type 1 transforms.

%fig:type2bottom
%(1)
For type 2 (lower plots in Fig.~\ref{fig:type12}),
the sensitivity to clustering is much weaker:
all libraries tackle ``cluster'' at about the same speed they tackle
``rand'', 
apart from {cuFINUFFT} which becomes 3--4$\times$ faster.
%except for large problem sizes ($N_i\geq 2^{10}$)
%(2)
While {cuFINUFFT} has similar
``total+mem'' time as {CUNFFT}, its ``exec'' time is 2--5$\times$ faster than that of {CUNFFT}.
In 3D our detailed findings are quite similar, and we do not show them.

%table:
Lastly, in Table \ref{table:3dtype1} we detail the type 1 performance and RAM usage of cuFINUFFT in 3D, for two tolerances. 
%(1)
We see again that higher speedup with respect to FINUFFT
is achieved for \textit{low accuracy} and \textit{large problem sizes}.
%(2)
The spreading method \textbf{SM} achieves better performance, but at a cost of
slightly more GPU RAM usage for large problems.
%(3)
Lastly, spreading is still the performance bottleneck for 3D type 1:
it occupies over $90\%$ of ``exec'' time for all
accuracies and problem sizes.

% DDDDDDDDDDDDDDDDDDDDDDDDDDDDDDDDDDDDDDDDDDDDDDDDDDDDDDDDDDDDDDDDDDDDDDDDDDDD
\paragraph{Double precision comparisons}
Fig. \ref{fig:double} compares the performance for both types in 2D and 3D,
for all libraries (except gpuNUFFT, whose $\epsilon$ appears always
to exceed $10^{-3}$).
% 2D type 1
We see from the top left plot that for 2D type 1, {cuFINUFFT}
outperforms the others by 1--2 orders of magnitude.
The best performance is achieved by \textbf{SM} (blue) for
high accuracy ($\epsilon\leq 10^{-5}$) and by \textbf{GM-sort} (cyan) for low
accuracy. The ``exec'' speedup of {cuFINUFFT} (taking the faster of
\textbf{SM} and \textbf{GM-sort}), vs {FINUFFT}, ranges from 4--11$\times$.
% 3D type 1
From the top right plot, for 3D type 1, {cuFINUFFT} is only faster than 
{FINUFFT} for relative error $\epsilon\geq 10^{-10}$, merely matching its speed
at the highest accuracies.
% 2,3D type 2
From the bottom plots, for type 2, {cuFINUFFT} is always the fastest,
and by a large factor at high accuracies.
The ``exec'' time of {cuFINUFFT} is on average 6$\times$ faster than that
{FINUFFT} in both dimensions.
In 2D, and low-accuracy 3D, we see that host-to-device transfer dominates:
a several-fold speedup is available by maintaining data on the GPU.

% MMMMMMMMMMMMMMMMMMMMMMMMMMMMMMMMMMMMMMMMMMMMMMMMMMMMMMMMMMMMMMMMMMMMMMMMMMMMM
%\section{Multi-GPU Applications on High-Performance Systems}
\section{Multi-GPU Applications}
\label{s:multi}

% The nascent exascale supercomputers are opening exciting new opportunities in
% the field of x-ray scattering imaging. One such application is the analysis of
% single-molecule scattering data from free electron lasers: an unknown sample is
% irradiated by a coherent x-ray source, and the molecular structure is
% reconstructed from the measured intensity pattern. This is a challenging from
% the perspective of model reconstruction, because many images (on the order of
% 100 GB to several TB) are required to reconstruct the molecular structure of
% the unknown sample. Furthermore, timely reconstruction is necessary as
% measurement time at free electron lasers is limited and expensive. Hence,
% efficient software libraries are required to process data at the exaflop scale.

\begin{figure}
    \centering
    \includegraphics[width=0.4\textwidth]{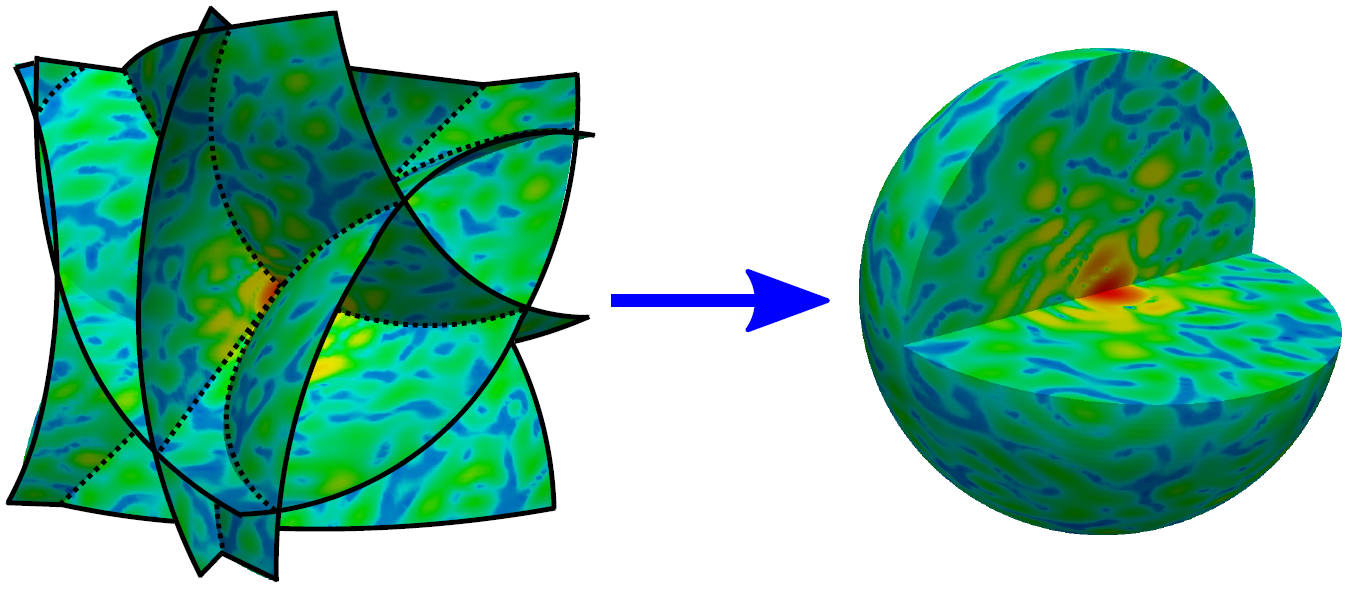}
    \caption{%
      M-TIP merging step: 3D Fourier transform data, collected on multiple
      Ewald sphere slices with arbitrary orientations, is merged onto a single
      uniform grid.  Image credit: Jeffrey Donatelli (Lawrence Berkeley
      National Laboratory).  \label{fig:mtip:merging}%
    }
\end{figure}

Finally, we illustrate the multi-GPU performance of cuFINUFFT in 3D coherent
X-ray image reconstruction.  Single particle imaging is a technique whereby the
3D electron density of a molecule may be recovered at sub-nanometer resolution from a
% small? *** D'17 paper says 24 images, but we talk about 10^5, so I'm confused.
% small was a few years ago, but now we're using large (10^5 image) data sets -- Johannes
large ($\le 10^5$) set of 2D far-field diffraction images, each taken in a
single shot of a free-electron laser with a random {\em unknown} molecular
orientation \cite{donatelli2017}.  Each 2D image measures the squared magnitude
Fourier transform of the density on an (Ewald sphere) slice passing through the
origin; see Fig.~\ref{fig:mtip:merging}.  The {\em multitiered iterative
phasing} (M-TIP) algorithm is used for reconstruction \cite{donatelli2017}.
Broadly speaking, one starts with a Fourier transform estimate on a 3D
Cartesian grid,
% *** or is it spherical polar? If not cart, how is a type 2 relevant?
% I think this is clear now, cf my email, right? -- Johannes
and estimated orientations, then iterates the following four steps:
\ben
\item[i)] ``Slicing'': a 3D type 2 NUFFT is used to evaluate the Fourier
  transform on a large set of Ewald sphere slices.
  % *** did I screw this up: is it only at the 24 or so estimated orientations?
  % no, this is between $10^3$ and $10^4$ -- Johannes
\item[ii)] 
  Orientation matching: adjust each slice orientation using its 2D image data.
  % *** and... did I screw this up? Not sure if match to a discrete but large set of slices evaluated in step i) ?
  % no screwup, this seems fine to me -- Johannes
\item[iii)] ``Merging'': solve for 3D Fourier transform data matching
  the 2D images on known slices, as in Fig.~\ref{fig:mtip:merging}; this needs
  two 3D type 1 NUFFTs.
\item[iv)] Phasing: find the most likely phase of the 3D Fourier transform
  to give a real-space density of known support.
\een
  
% In order to achieve exaflop performance, an implementation of the M-TIP
% algorithm needs to offload some of the work to hardware accelerators -- such as GPUs.
The code acceleration strategy, a part of the Exascale Computing Project, has
been to offload the intensive steps i)--iii) to GPUs.
%as those scale with the size of the input data.

\subsection{Work management and the {cuFINUFFT} Python interface}

We use MPI to manage parallel processes, via the mpi4py package. Each
MPI {\em rank} (i.e.\ process) is assigned some data to handle and a GPU.
%(there may be more than one rank on a given GPU).  - J.A. was confused about.
Since slicing and merging are linear operations, we can scatter
(\texttt{mpi4py.scatter}) before the slicing step, and reduce
(\texttt{mpi4py.reduce}) after the merging step.  In modern HPC environments
each compute node is furnished with several GPUs---e.g. NERSC's Cori GPU system
has 8 NVIDIA V100 per node, while OLCF's Summit system has 6 V100 per node.
Thus, depending on the ratio of GPUs to CPU cores, we can have several MPI
ranks share the same GPU.  For M-TIP, load balancing is simple because each
rank does (roughly) the same amount of work, thus we assign GPUs in a
round-robin fashion. We use PyCUDA to transfer data between device and
host. This allows cuFINUFFT to access \texttt{numpy.ndarray} objects as
\texttt{double []}, hence requiring no specialized API calls to convert data
between Python and cuFINUFFT. In order to ensure that the PyCUDA API
sends the data to the correct device, we manually define the device context.
Here is an example of taking a type 1 3D NUFFT
with nonuniform coordinates
{\tt X}, {\tt Y}, {\tt Z}, and strengths {\tt nuvect}
(note: mpi4py provides \texttt{rank}):

\begin{python}[language=python]
from cufinufft import cufinufft
from pycuda.gpuarray import GPUArray, to_gpu

# Initialize GPU using round-robin assignment
GPUS_PER_NODE = 8              # Cori GPU
device_id = rank % GPUS_PER_NODE

pycuda.driver.init()
device = pycuda.driver.Device(device_id)
ctx = device.make_context()

ugrid_gpu = GPUArray(shape)    # Memory for result

plan = cufinufft(1, shape, eps=eps,
                 gpu_device_id=device_id)
plan.set_pts(to_gpu(X), to_gpu(Y), to_gpu(Z))

plan.execute(to_gpu(nuvect), ugrid_gpu)
\end{python}

The cuFINUFFT Python interface allows the user to assign a {\tt cufinufft} plan
to a specific device by setting the {\tt gpu\_device\_id} option.
See the documentation on GitHub for details.
%The {\tt cufinufft.default\_opts} function selects device 0.
The M-TIP code requires a tolerance of $\varepsilon=10^{-12}$.

% tttttttttttttttttttttttttttttttttttttttttttttttttttttttttttttttttttttttt
\begin{table*}
    \begin{center}
        \begin{tabular}{c c c c c c c c}
			\toprule
            Task                  &
            Uniform grid (per rank) &
            Nonuniform points (per rank) &
            Density           &
            Parallelism                &
            CPU time [s]              &
            GPU time [s]              &
            GPU time [s]
            \\
                              &
            $N_1=N_2=N_3$             &
            $M$               &
            $\rho$ &
                              &
            (Intel Skylake)   &
            (Cori GPU)        &
            (Summit)
            \\
            \midrule
            Slicing & $41$ & $1.02\times 10^6$ & $1.86$ & single-rank & $0.11$ & $0.075\quad(1.5\times)$ & $0.076\quad(1.5\times)$
            \\
            (type 2) & & & & whole-node & $0.95$ & $0.078\quad(12\times)$ & $0.11\quad(8.6\times)$
            \\
            \midrule
            % Slicing & $(81, 81, 81)$ & $32768000$ & $\rho = 61.7$ & $6062$ & $1707$ & --
            % \\
            % Large Problem & & & & & $(3.55\times)$ &
            % \\
            % \hline
            Merging & $81$ & $1.64\times 10^7$ & $3.85$ & single-rank & $1.62$ & $1.89\quad(0.9\times)$ & $1.76\quad(0.9\times)$
            \\
            (type 1) & & & & whole-node & $9.97$ & $1.94\quad(5.1\times)$ & $1.76\quad(5.7\times)$
            \\
            %\hline
            % Merging & $(81, 81, 81)\quad -$ & $16384000$ & $\rho = 3.85 \quad -$ & $2465$ & $1932$ & --
            % \\
            % Large Problem & $(162, 162, 162)$ & & $30.8$ & & $(1.3\times)$ &
            % \\
            % \hline
            \bottomrule
        \end{tabular}
        \caption{%
            Problem sizes and average NUFFT wallclock times for a representative
            M-TIP iteration: NERSC's Cori GPU (Intel Skylake with $8$ NVIDIA
            V100's) and OLCF's Summit (IBM Power9 with $6$ NVIDIA V100's).  The
            problem sizes are fixed per MPI rank. The CPU code is FINUFFT v1.1.2, with 40
            threads on the Intel Skylake. The speedup ratio of single-
            or multi-GPU cuFINUFFT over the CPU code is shown in parentheses.
            Rows labeled ``whole-node'' use problems scaled up by the
            number of GPUs per node---8 for Cori GPU
            %(and Intel Skylake)
            and 6 for Summit---using this same number of ranks (i.e., one rank per GPU).
            \label{tab:mtip}%
        }
    \end{center}
\end{table*}

\subsection{cuFINUFFT Performance on Cori GPU and Summit}

To perform a single-node weak scaling study we
used, per rank, the NUFFT problem sizes in Table~\ref{tab:mtip},
which correspond to $10^3$ images.
%Its goal is to determine the optimal proportion of MPI ranks to
%GPUs possible on Cori GPU and Summit.
% *** I don't get this goal.
%
% A small
% problem is typically of the oder of $10^4$ images -- these are broken up into
% batches of $10^3$, with a rank processing a single batch. Depending on how many
% ranks run on a single compute node, anywhere between $10^3$ to $4\times 10^4$
% images are processed by a single compute nodes. Hence a ``small'' job runs on 1
% -- 4 compute nodes.
% Large problems -- ones which more closely represent a
% ``realistic'' data set will consist of over $5\times 10^5$ images, these are
% divided into batches of $2\times 10^3$ and processed on over 12 compute nodes
% -- depending on actual problem size. Table~\ref{tab:mtip} shows the problem
% sizes \emph{per rank} for a typical small and large problem.
%
The table shows the average wall-clock time spent performing
NUFFTs during slicing (type 2 NUFFT) and merging (type 1 NUFFT) steps for one
M-TIP iteration.  Comparing the GPU wall-clock times (including data movement)
to the equivalent problem running on 40 CPU threads on a single Intel Skylake Cori
GPU node (using the FINUFFT code), we find that cuFINUFFT on a single GPU is
roughly similar to the CPU times, while for
the larger problem distributed over the
whole node (multi-GPU) it is 5--12$\times$ faster than on the CPU.
%Therefore the potential
%for speedup is much greater following host-to-device data transfer
%optimizations.    *** Ahb doesn't agree, because at 1e-12 the NUFFTs dominate.
%
%Indeed, we have confirmed this by using NVIDIA NSight Systems to
%profile the entire launch of cuFINUFFT starting with the memory transfer
%from Python.
%  *** is a detailed point that readers won't get - I don't.

\begin{figure}
    \centering
    \includegraphics[width=0.48\textwidth]{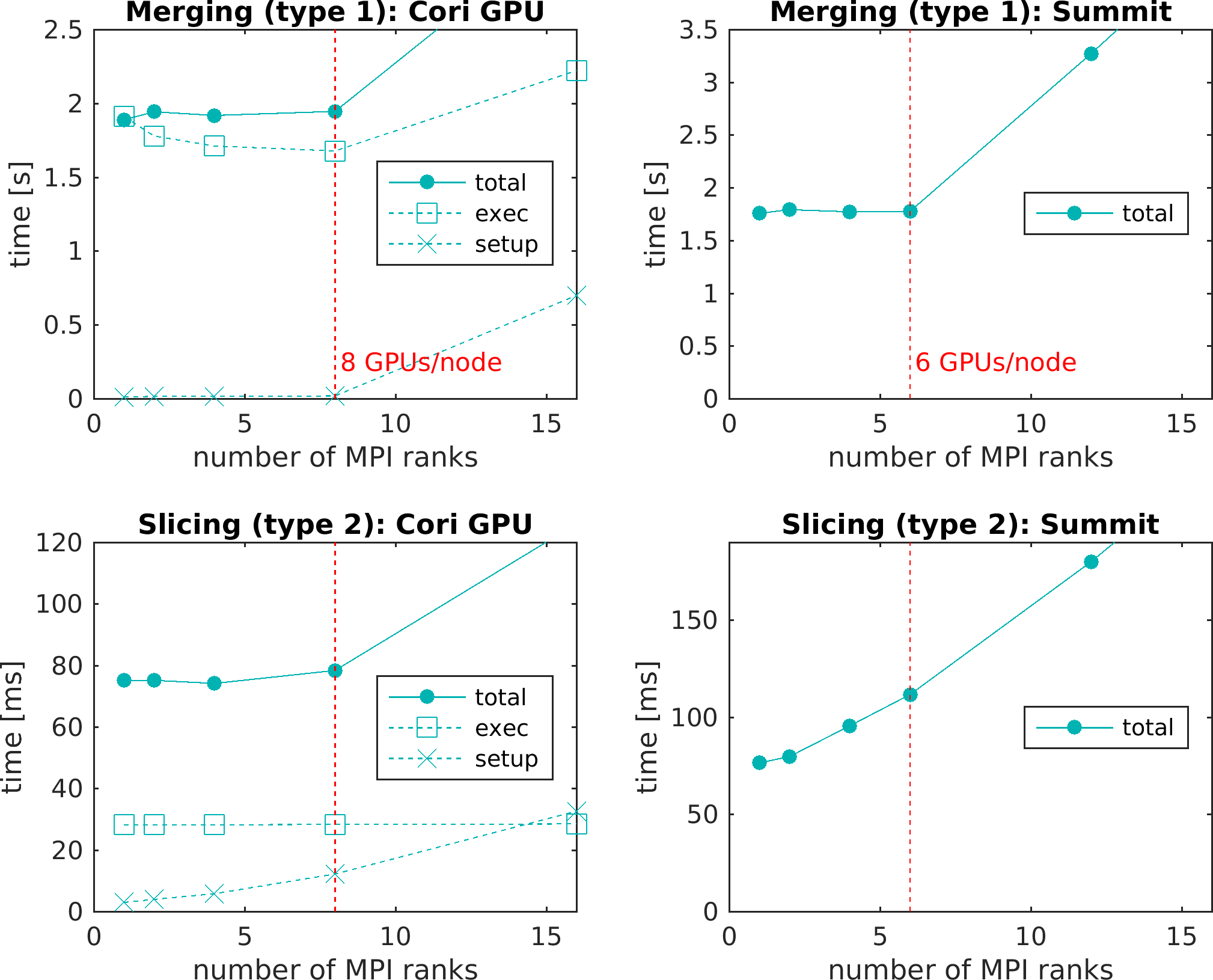}
    \caption{%
        Single-node multi-GPU weak scaling on NERSC Cori GPU (left) and OLCF
        Summit (right).
        %Vertical dotted lines show the number of GPUs/node.
        We achieve close to ideal weak scaling (flat lines) up to a number
        of ranks matching the number of GPUs on the node (vertical dotted line).
        \label{fig:mtip:scaling}%
    }
\end{figure}

Fig.~\ref{fig:mtip:scaling} shows the weak scaling performance on single
nodes of NERSC's Cori GPU and OLCF's Summit.  Each rank is given the
single-rank problem size from Table~\ref{tab:mtip}.  Solid lines show the total
time including host-device transfer.  Crosses show setup time (plan, input and
nonuniform point sorting), while squares show NUFFT execution time.  In all
cases (except type 2 on Summit), we see ideal weak scaling up to one rank per
GPU.
We found that enabling multi-process service (MPS) made no measurable
difference.
% see: https://docs.nvidia.com/deploy/pdf/CUDA_Multi_Process_Service_Overview.pdf
We see rapid deterioration of weak scaling once each GPU is used by more than
one rank, suggesting that cuFINUFFT uses each GPU to capacity.
%shows that latency hiding cannot be achieved
%merely by adding ranks. Indeed, the consistent execution time, yet diverging
%wallclock time indicates that this application is limited by the host-to-device
%memory transfer *** EDIT THAT.
% *** again, I don't agree.

% CCCCCCCCCCCCCCCCCCCCCCCCCCCCCCCCCCCCCCCCCCCCCCCCCCCCCCCCCCCCCCCCCCCCCCCCCCCC
\section{Conclusions}

We presented a general-purpose GPU-based library for nonuniform fast Fourier transforms: cuFINUFFT.
It supports both transforms of type 1 (nonuniform to uniform) and type 2 (uniform to nonuniform), in 2D and 3D, with adjustable accuracies.
%single/double precision.
%
By using an efficient kernel function,
sorting the nonuniform points into bins, and leveraging shared memory to reduce write collisions, cuFINUFFT obtains a significant speedup compared to established CPU- and GPU-based libraries.
%
%At low accuracies, our mean transform cost per new nonuniform point is
%1--3 ns, and when reusing points, sub-nanosecond.
%
On average, we observe a speedup of one order of magnitude over the FINUFFT parallel CPU library.
We also observe up to an order of magnitude speedup compared to the CUNFFT GPU library,
and more in the case of clustered type 1 transforms.
We also see excellent multi-GPU weak scaling in an
iterative 3D X-ray reconstruction application.

There are several directions for future work.
One is extending the library to include 1D and type 3 transforms, but also to use
smaller upsampling factors $\sigma$, which can significantly reduce memory size.
It is also worth exploring supporting other GPUs via a library that provides a unified hardware API, such as OCCA
%\cite{medina2014occa}
or Kokkos.

\section{Acknowledgments}

This research used resources of two U.S. Department of Energy Office of Science
User Facilities: NERSC and OLCF (contract \# DE-AC02-05CH11231 and
DE-AC05-00OR22725). Johannes Blaschke's work is supported by the DOE OOS in part through DOE's ECP ExaFEL project, Project \# 17-SC-20-SC, a
collaborative effort of
%two DOE organizations
the Office of Science and the
National Nuclear Security Administration. We thank Jeff Donatelli, Chuck Yoon,
and Christine Sweeney for work on the M-TIP code, and 
Georg Stadler
% from New York University  - unusual to state this, don't need.
for helpful suggestions at various
stages of this work.  
The work was conducted while Joakim And\'en was a visiting scholar at CCM.
%Center for Computational Mathematics of the
%in the Flatiron Intitute.
Part of Yu-hsuan Shih's work was supported through the CCM internship program.
The Flatiron Institute is a division of the Simons Foundation.
% Per Amit Singer
The work of Garrett Wright was supported by the Moore Foundation Award \#9121 Cryo-EM Software Grant.

% BBBBBBBBBBBBBBBBBBBBBBBBBBBBBBBBBBBBBBBBBBBBBBBBBBBBBBBBBBBBBBBBBBBBBBBBBB
\bibliographystyle{unsrt}
\bibliography{references}
\end{document}